\documentclass{article}

\usepackage{arxiv}
\newcommand{\undertitle}[1]{}

\usepackage[utf8]{inputenc} 
\usepackage[T1]{fontenc}    

\usepackage{url}            
\usepackage{booktabs}       
\usepackage{amsfonts}       
\usepackage{nicefrac}       
\usepackage{microtype}      
\usepackage{lipsum}		
\usepackage{graphicx}
\usepackage{natbib}
\usepackage{doi}
\usepackage[version=4]{mhchem}
\usepackage{caption}
\usepackage{float}
\usepackage[english]{babel}
\usepackage{pdftexcmds}
\title{Mid-Infrared Imaging Spectroscopy of N$_2$O Solid Simulating the haze of trans-Neptunian objects}
\usepackage{fontawesome5}
\usepackage{xcolor} 

\author{
Daiki Takama\textsuperscript{1*}~\href{https://orcid.org/my-orcid?orcid=0009-0000-6007-1634}{\textcolor[HTML]{A6CE39}{\faOrcid}}\\[2mm]
Ryoichi Koga\textsuperscript{2}~\href{https://orcid.org/0000-0002-4278-7804}{\textcolor[HTML]{A6CE39}{\faOrcid}}\\[2mm]
Shohei Negishi\textsuperscript{3}~\href{https://orcid.org/0009-0008-2900-2954}{\textcolor[HTML]{A6CE39}{\faOrcid}}, Biao Zhao\textsuperscript{3}~\href{https://orcid.org/0009-0000-0861-9527}{\textcolor[HTML]{A6CE39}{\faOrcid}}, Yuan Li\textsuperscript{3}~\href{https://orcid.org/0009-0009-5594-5160}{\textcolor[HTML]{A6CE39}{\faOrcid}},Yasuhiro Hirahara\textsuperscript{3}~\href{https://orcid.org/0000-0003-2776-1849}{\textcolor[HTML]{A6CE39}{\faOrcid}}\\[2mm]
Fumiyuki Ito\textsuperscript{4}~\href{https://orcid.org/0000-0002-6654-5172}{\textcolor[HTML]{A6CE39}{\faOrcid}}
}

\hypersetup{
pdftitle={A template for the arxiv style},
pdfsubject={q-bio.NC, q-bio.QM},
pdfauthor={David S.~Hippocampus, Elias D.~Striatum},
pdfkeywords={First keyword, Second keyword, More},
}

\begin{document}
\maketitle

\begin{abstract}

\ Nitrous oxide (N$_2$O) ice is likely to exist in trans-Neptunian objects such as Pluto and Triton, potentially formed through ultraviolet (UV) radiation from the Sun or cosmic ray irradiation of N$_2$ and CO ices. However, the mid-infrared spectral characteristics of N$_2$O ice in higher temperature regions (90-110 K), changes in mid-infrared spectra during UV irradiation, and the chemical network of nitrogen oxide (N$_x$O$_y$) ices remain insufficiently understood. This study aims to elucidate these aspects through in-situ mid-infrared spectral measurements of cryogenic particles using two-dimensional imaging Fourier transform infrared spectroscopy.

Spectroscopic imaging confirmed strong absorption at 7.75 {\textmu}m (N$_2$O $\nu_1$ vibrational mode), with weaker vibrational modes observed at 8.60 {\textmu}m (N$_2$O 2$\nu_2$), 7.27 {\textmu}m (N$_2$O torsion), and 5.29 {\textmu}m (N$_2$O $\nu_1$+$\nu_2$). Annealing experiments simulating high-temperature conditions demonstrated that all vibrational modes irreversibly intensified with increasing temperature, indicating progressive crystallization. New spectral features appeared at approximately 12 {\textmu}m and 14 {\textmu}m at the condensed sample.

N$_2$O ice was exposed to ultraviolet radiation (190-340 nm) using a D$_2$ lamp for 8.5 hours to investigate spectral changes during UV irradiation. After 60-90 minutes of irradiation, all N$_2$O vibrational modes disappeared, while absorption intensities of various nitrogen oxides, including NO, NO$_2$, N$_2$O$_3$, and O$_3$ increased. Beyond 180 minutes, vibrational modes of multiple nitrogen oxide ices exhibited intensity variations across different wavelengths, corresponding to other species such as cis-(NO)$_2$, N$_2$O$_4$, and N$_2$O$_5$.

\end{abstract}


\maketitle

\clearpage
\section{Background}
\ Nitrous oxide (N$_2$O) is a crucial substance for generating life precursor amino acids, with its solid form at low temperatures creating a molecular crystal through strong electrostatic interactions due to significant molecular polarization. These prebiotic amino acids form through the dissociation and recombination of nitrogen compounds, with N$_2$O functioning as an essential seed material in this formation process (Bergantini et al. \protect\hyperlink{cite.Bergantini2022a}{2022a}).

\ Despite extensive spectroscopic observations of various planetary surfaces, atmospheres, and interstellar ice dust (Boduch et al. \protect\hyperlink{cite.Boduch2015}{2015}; Allodi et al. \protect\hyperlink{cite.Allodi2013}{2013}), infrared spectra of N$_2$O ice remain unreported. The vibrational spectra of N$_2$O ice in the infrared region likely exhibit complex responses due to solid particle growth, metamorphism, and photodissociation processes triggered by solar ultraviolet (UV) radiation.

Previous observations by planetary exploration satellites Voyager 2 and New Horizons have revealed geologically active cryovolcanoes on Pluto and Triton (Grundy et al. \protect\hyperlink{cite.Grundy2016}{2016}). Portions of these surfaces feature geysers emitting nitrogen compounds (Schenk et al. \protect\hyperlink{cite.Schenk2021}{2021}; Singer et al. \protect\hyperlink{cite.Singer2022}{2022}). Research suggests that N$_2$O ice forms on the surfaces and in the atmospheres of Pluto and Triton when N$_2$ interacts with CO, CO$_2$, and O$_2$ under exposure to UV radiation and cosmic rays (Equation 1) (Jamieson et al. \protect\hyperlink{cite.Jamieson2005}{2005}; Almeida et al. \protect\hyperlink{cite.Almeida2017}{2017}).

\begin{equation}
\ce{N2 + O(^1D) -> N2O(X^1\Sigma^+)} \tag{1}
\end{equation}

\ As shown in Figure 1, Voyager 2, New Horizons, and ALMA telescope data have enabled predictions of temperature and pressure distributions relative to altitude, as well as number densities of N$_2$ ice on Pluto and Triton. Strobel and Zhu \protect\hyperlink{cite.Strobel2017}{2017} predicted these celestial bodies maintain tenuous atmospheres with temperatures ranging from approximately 38 K (surface) to 110 K (at altitudes of several tens of km) and pressures between $10^{-1}$ and $10^{-2}$ Pa. Previous studies have attempted environmental simulation spectroscopic observations of N$_2$O ice in the infrared spectrum.

\ Mifsud et al. \protect\hyperlink{cite.Mifsud2022}{2022} irradiated N$_2$O ice at 20 K and $10^{-4}$ Pa with 5 keV electrons simulating cosmic radiation, resulting in the formation of N$_2$O$_4$ and O$_3$ ice in both crystalline and amorphous phases. They observed different intensity ratios between the $\nu_1$ and 2$\nu_2$ modes in crystalline versus amorphous phases, with the amorphous phase exhibiting lower intensities than the crystalline phase.

Barros et al. \protect\hyperlink{cite.deBarros2017}{2017} measured transmission absorption spectra of N$_2$O ice between 10 K and 80 K in the mid-infrared wavelength region corresponding to fundamental vibrational absorptions of N$_2$O molecules. From 10 K to 40 K, absorbance of $\nu_1$ (symmetric stretching), 2$\nu_2$ (antisymmetric stretching), and $\nu_3$ (bending) vibrations increased as crystallization progressed. Subsequently, these vibrations showed decreased absorbance from 40 K to 80 K as the sublimation curve was exceeded.

Hudson et al. \protect\hyperlink{cite.Hudson2017}{2017} first measured band strengths and absorption coefficients of amorphous and crystalline N$_2$O ice. They emphasized that temperature and deposition rate are crucial in N$_2$O ice formation. Increasing deposition thickness from 0.04 to 0.16 {\textmu}m at 10 K and $10^{-6}$ Pa revealed clearer spectral structures of amorphous phase vibrations: $\nu_3$ (bending) at ~4.47 {\textmu}m, $\nu_1$ (symmetric stretching) at ~7.73 {\textmu}m, and $\nu_2$ (antisymmetric stretching) at ~16.9 {\textmu}m. Maintaining a constant deposition thickness of 0.18 {\textmu}m while warming from 10 K to 70 K induced a phase transition from amorphous to crystalline between 24 K and 37 K.

\ These various environmental simulation spectroscopy experiments have yielded diverse insights into N$_2$O ice formation processes and behaviors. However, since New Horizons observations (Singer et al. \protect\hyperlink{cite.Singer2022}{2022}), no mid-infrared spectroscopic observations have simulated high-temperature regions above 10 km altitude (90 K - 110 K) where materials erupt from cryovolcanoes on Pluto and Triton. Furthermore, the relationship between chemical evolution processes induced by UV radiation and vibrational spectra in the infrared region remains unclear.

\ One approach involves infrared spectroscopy experiments on N$_2$O ice simulating Pluto and Triton's surface and atmospheric conditions. While obtaining solid-state information in the radio region corresponding to rotational transitions is theoretically impossible due to constrained molecular rotation in solids, the infrared region corresponding to vibration-rotation transitions allows simultaneous acquisition of gas and solid information. Our study represents the first attempt to obtain mid-infrared spectroscopic data in higher temperature regions and to clarify spectral changes induced by simulated solar UV radiation, offering significant scientific value. Predicting N$_2$O ice chemical networks under solar UV radiation will also facilitate comparisons with future planetary exploration satellite observations.

\begin{figure}[htbp]
  \centering
  \includegraphics[width=0.8\textwidth]{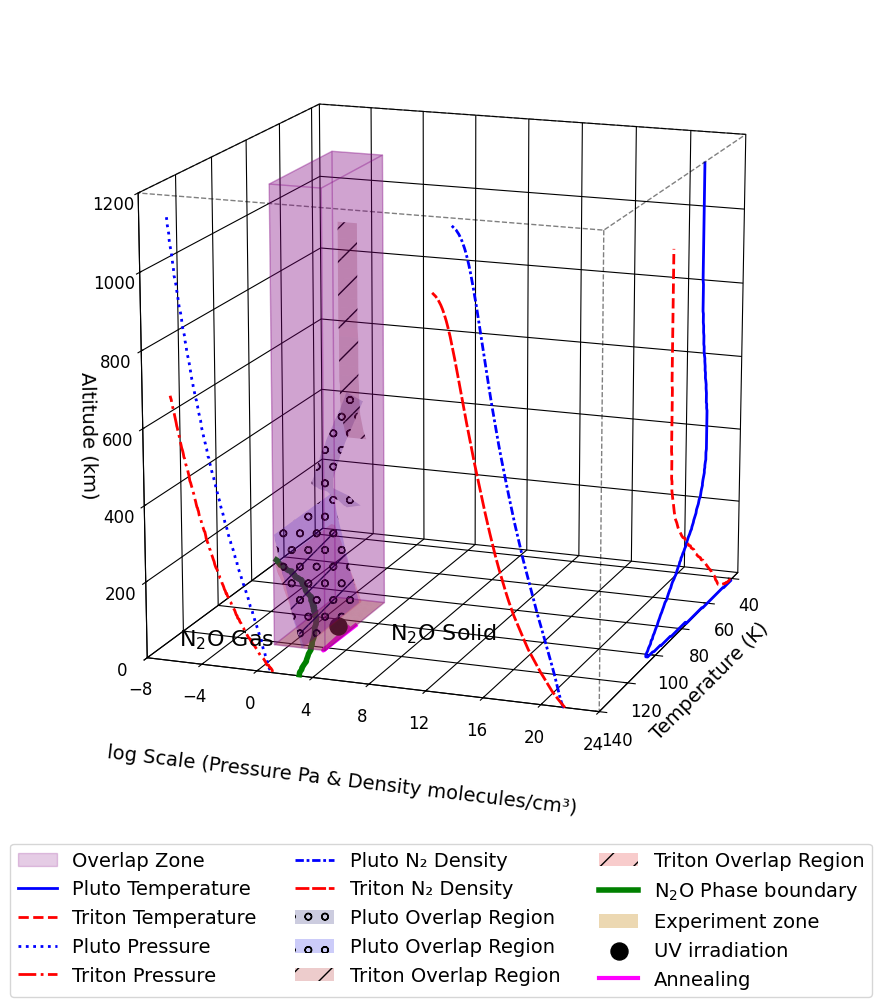}
  \caption{Temperature, pressure, and nitrogen ice number density distributions for altitude for Pluto and Triton (Strobel and Zhu \protect\hyperlink{cite.Strobel2017}{2017}). The pink region indicates the temperature and pressure conditions simulated in this study. The green line on the bottom surface represents the phase boundary of N$_2$O (Ferreira and Lobo \protect\hyperlink{cite.Ferreira2009}{2009}). The temperature and pressure range during annealing and UV irradiation in this experiment are indicated by pink lines and black dots. Additionally, each celestial body's intersection regions of pressure and temperature are represented as dotted areas within the pink region (Overlap Region).}
  \label{fig:3D}
\end{figure}

\clearpage
\section{Methods}
\subsection{Experimental equipment}
\ The 2D FT-MIR operates based on a quasi-common-path wavefront-division phase-shifting interferometry method (Qi et al. \protect\hyperlink{cite.Qi2015}{2015}), which introduces spatial phase differences through wavefront division and amplifies interference light components at the imaging plane for detection.

A mid-infrared light source (SLS303, Thorlabs Co., $0.55\,\mathrm{\mu m} < \lambda < 15\,\mathrm{\mu m}$), a cryostat with vacuum chamber, and a 2D FT-MIR spectrometer (NK-0812-TD-NU, Nisshin Machinery Co., wavelength range $8\,\mathrm{\mu m} < \lambda < 12\,\mathrm{\mu m}$, wavelength resolution $\Delta\lambda$ 0.5 {\textmu}m) are aligned linearly to measure the growth and metamorphism processes of N$_2$O ice through mid-infrared transmission absorption imaging spectroscopy (Figure 2) (Koga et al. \protect\hyperlink{cite.Koga2024}{2024}).

The mid-infrared light source irradiates the sample holder perpendicularly with mid-infrared radiation. ZnSe aspheric lenses serve as the incident and exit windows on the vacuum chamber walls, with focal lengths of 63.5 mm and 25 mm, respectively. By setting the distance from the sample holder to the exit ZnSe lens at 52.5 mm and the distance from the exit ZnSe lens to the 2D FT-MIR multi-slit at 44.5 mm, the N$_2$O sample can be observed at unity magnification on the 2D FT-MIR detector.

\begin{figure}[htbp]
  \centering
  \includegraphics[width=0.8\textwidth]{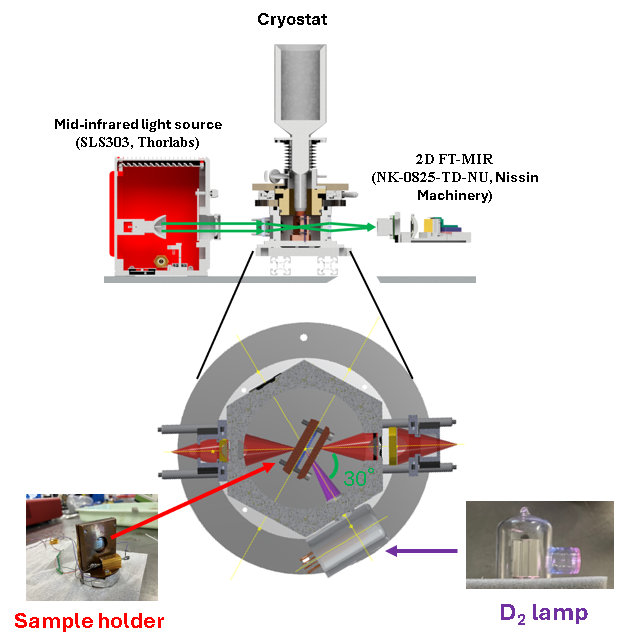}
  \caption{Experimental setup with the mid-infrared light source, vacuum chamber-equipped cryostat, and 2D FT-MIR arranged in a straight line for spectroscopic measurement of N$_2$O ice growth and alteration processes via mid-infrared transmission absorption imaging spectroscopy. N$_2$O samples are deposited on a ZnSe plate-equipped sample holder. The distance from the sample holder to the ZnSe lens is set to 52.5 mm, the distance from the ZnSe lens to the multi-slit of the 2D FT-MIR is set to 44.5 mm, and the N$_2$O sample is observed on the 2D FT-MIR detector at a scale of 1:1. A 39 $\Omega$ metal-clad variable resistor is fixed to the top of the sample holder using aluminum tape using the thermal anchor method. In addition, a D$_2$ lamp (manufactured by Mitrika Co.) is installed at about 30$^{\circ}$ to the deposited N$_2$O ice.}
  \label{fig:cryostat}
\end{figure}

\subsection{Procedure}
\ Initially, the vacuum chamber is purged with nitrogen, followed by approximately one hour of baking. This baking process prevents corrosion inside the vacuum chamber by removing water, carbon dioxide, and compounds such as nitric acid. Residual gases in the vacuum chamber without deposition and gases generated from sublimated N$_2$O solid samples are evacuated through a vacuum bellow rough pumping port equipped with a liquid nitrogen trap in the release path. This evacuation utilizes a pump with corrosion-resistant oil (PD-52, ULVAC).

\ Subsequently, the main vacuuming port valve on the vacuum chamber wall is opened. A backing dry scroll pump (IDP-3, Agilent Technologies) operates for approximately 15 minutes, followed by a main vacuuming dry turbo molecular pump (Agilent TwisTorr 74 FS, Agilent Technologies) running for about 1 hour to depressurize the vacuum chamber. Next, a few drops of ethanol are added to the interior of the Dewar bottle before filling it with liquid nitrogen to cool the sample holder with a ZnSe plate connected to the cold head. Aluminum tape is applied to a ~2 mm wide groove on the upper part of the sample holder, employing the Thermal anchor method to reduce heat influx from the external environment. Silver paste is applied to the contact area of the cold head, and aluminum retaining screws are used to enhance thermal conductivity. The sample holder temperature is reduced to approximately 85 K, and the pressure inside the vacuum chamber is decreased to $10^{-2}$$\sim10^{-3}$ Pa.

\ Next, N$_2$O standard gas (99.9 \% purity, 10 L, Taiyo Nissan Co.) from a gas cylinder is ejected through a pulse nozzle (009-0181 900, Parker Hannifin) at an angle of approximately 30$^{\circ}$ onto the ZnSe plate inside the vacuum chamber, with the pressing pressure regulated at 0.32 MPa using a regulator (NHW-1S-11NTF). During this process, a pulse nozzle driver (Multi-Channel IOTA ONE, Parker Hannifin) sets the nozzle opening time (On-time) and closing time (Off-time) to 100 ms and 900 ms, respectively.

\ Subsequently, spectral reference data are acquired through spectroscopic imaging immediately before ejecting the N$_2$O gas. After N$_2$O gas ejection, the sample holder temperature is maintained at 100 K for 20 minutes to allow N$_2$O ice growth, and spectra are obtained.

\ Next, annealing is performed by varying the temperature between 104 K and 122 K through heating and cooling processes. This temperature control utilizes a temperature controller connected to a 39 $\Omega$ metal-clad variable resistor at the voltage output terminal inside the vacuum chamber, achieving a $\Delta$1 K precision.

\ In a separate process, after growing N$_2$O ice, a D$_2$ lamp (manufactured by Mitrika Co.) with wavelengths of 190 nm - 340 nm irradiates the sample at an angle of approximately 30$^{\circ}$ for several hours at a temperature of 107 K.

\clearpage
\section{Results}

\subsection{Transmission absorption spectra of N\texorpdfstring{$_2$}{2}O ice and N\texorpdfstring{$_2$}{2}O gas}
\ Figure 3 presents two-dimensional transmission absorption spectra when N$_2$O gas was ejected with identical pulse parameters into the vacuum chamber under 107 K cooling (red) and at 302.4 K room temperature (blue). The spectra are mapped in 11$\times$11 grids, with each grid's spectrum representing an average of a 30$\times$30 pixel area, corresponding to 360$\times$360 {\textmu}m on the 2D FT-MIR detector. Each grid's horizontal axis spans 5-15 {\textmu}m, while the vertical axis represents absorbance from -0.5 to 1.5. Note that the wavelength ranges $5\,\mathrm{\mu m} < \lambda < 7\,\mathrm{\mu m}$ and $14\,\mathrm{\mu m} < \lambda < 15\,\mathrm{\mu m}$ exhibit significant noise as they fall outside the manufacturer's guaranteed range ($8\,\mathrm{\mu m} < \lambda < 12\,\mathrm{\mu m}$). Figure 4 shows an enlarged view of transmission absorption spectra from 4 grids near the center of the N$_2$O condensed sample.

\ Under 107 K cooling, several absorption bands are observed: a strong band at 7.75 {\textmu}m corresponding to the $\nu_1$ vibrational mode (N=O symmetric stretching vibration) of N$_2$O molecules, a band at 8.60 {\textmu}m for the 2$\nu_2$ vibrational mode (antisymmetric stretching), a torsion mode at 7.27 {\textmu}m (Dows \protect\hyperlink{cite.Dows1957}{1957}), a weak absorption band of the torsion mode at 7.96 {\textmu}m, and an absorption band of the $\nu_1$+2$\nu_2$ mode at 5.29 {\textmu}m. All five bands exhibit singlet structures (Table 1). This suggests that N$_2$O molecules exist in a condensed phase (N$_2$O ice) due to restricted rotational motion. In contrast, at 302.4 K room temperature, P and R branch doublet structures resulting from vibration-rotation interactions of the $\nu_1$ and 2$\nu_2$ vibrational modes are confirmed, indicating that N$_2$O molecules exist as gas (N$_2$O gas). Additionally, the absorption peaks of the $\nu_1$ vibrational mode, 2$\nu_2$ vibrational mode, and torsion mode appear stronger in N$_2$O ice than in N$_2$O gas. This suggests that N$_2$O molecules in the condensed phase, with bond dipole moments of $\frac{\partial\mu}{\partial\mu_{\text{N-N}}} = 2.45\,\text{D}/\text{\AA}$ and $\frac{\partial\mu}{\partial\mu_{\text{N-O}}} = 5.13\,\text{D}/\text{\AA}$, form a molecular crystal through strong electrostatic interactions due to significant polarization. These polarized N$_2$O molecules enhance intermolecular interactions via van der Waals forces in the condensed phase.

\begin{table}[ht]
\centering
\captionsetup{justification=centering}
\caption{N$_2$O absorption bands}
\begin{tabular}{p{4.5cm}p{4.5cm}p{5cm}}
\hline
Chemical species & Wavelength/\textmu m & Vibrational mode \\
\hline
N$_2$O (ice) & 7.75 & $\nu_1$ \\
N$_2$O (ice) & 8.60 & 2$\nu_2$ \\
N$_2$O (ice) & 7.27 & torsion \\
N$_2$O (ice) & 5.29 & $\nu_1$+$\nu_2$ \\
N$_2$O (ice) & 7.96 & 2$\nu_2$+91 (torsion) \\
N$_2$O (gas) & 7.70/7.84 & $\nu_1$ (R-branch/P-branch) \\
N$_2$O (gas) & 8.52/8.75 & 2$\nu_2$ (R-branch/P-branch) \\
\hline
\end{tabular}
\label{tab:1}
\end{table}

\begin{table}[ht]
\centering
\captionsetup{justification=centering}
\caption{N$_2$O absorption bands (comparison with previous studies)}
\begin{tabular}{p{4cm}p{3cm}p{3cm}p{4cm}}
\hline
Chemical species & Wavelength/\textmu m & Vibrational mode & Reference \\
\hline
N$_2$O (amorphous) & 7.79 & $\nu_1$ & Mifsud et al. \protect\hyperlink{cite.Mifsud2022}{2022}\\
N$_2$O (crystalline) & 7.79 & $\nu_1$ & // \\
$^{15}$N$^{14}$N$^{16}$O (crystalline) & 7.82 & $\nu_1$ & Hudson et al. \protect\hyperlink{cite.Hudson2017}{2017} \\
$^{14}$N$^{14}$N$^{18}$O (crystalline) & 7.96 & $\nu_1$ & // \\
N$_2$O (crystalline) & 8.58 & 2$\nu_2$ & // \\
N$_2$O (crystalline) & 5.29 & $\nu_1$+$\nu_2$ & // \\
N$_2$O (ice) & 7.73 & $\nu_1$ & Dows \protect\hyperlink{cite.Dows1957}{1957}
\\
N$_2$O (ice) & 8.57 & 2$\nu_2$ & // \\
N$_2$O (ice) & 7.22 & $\nu_1$+92 (torsion) & // \\
N$_2$O (ice) & 5.29 & $\nu_1$+$\nu_2$ & // \\
N$_2$O (ice) & 7.96 & 2$\nu_2$+91 (torsion) & // \\
\hline
\end{tabular}
\label{tab:2}
\end{table}

\begin{figure}[htbp]
  \centering
  \includegraphics[width=1.0\textwidth]{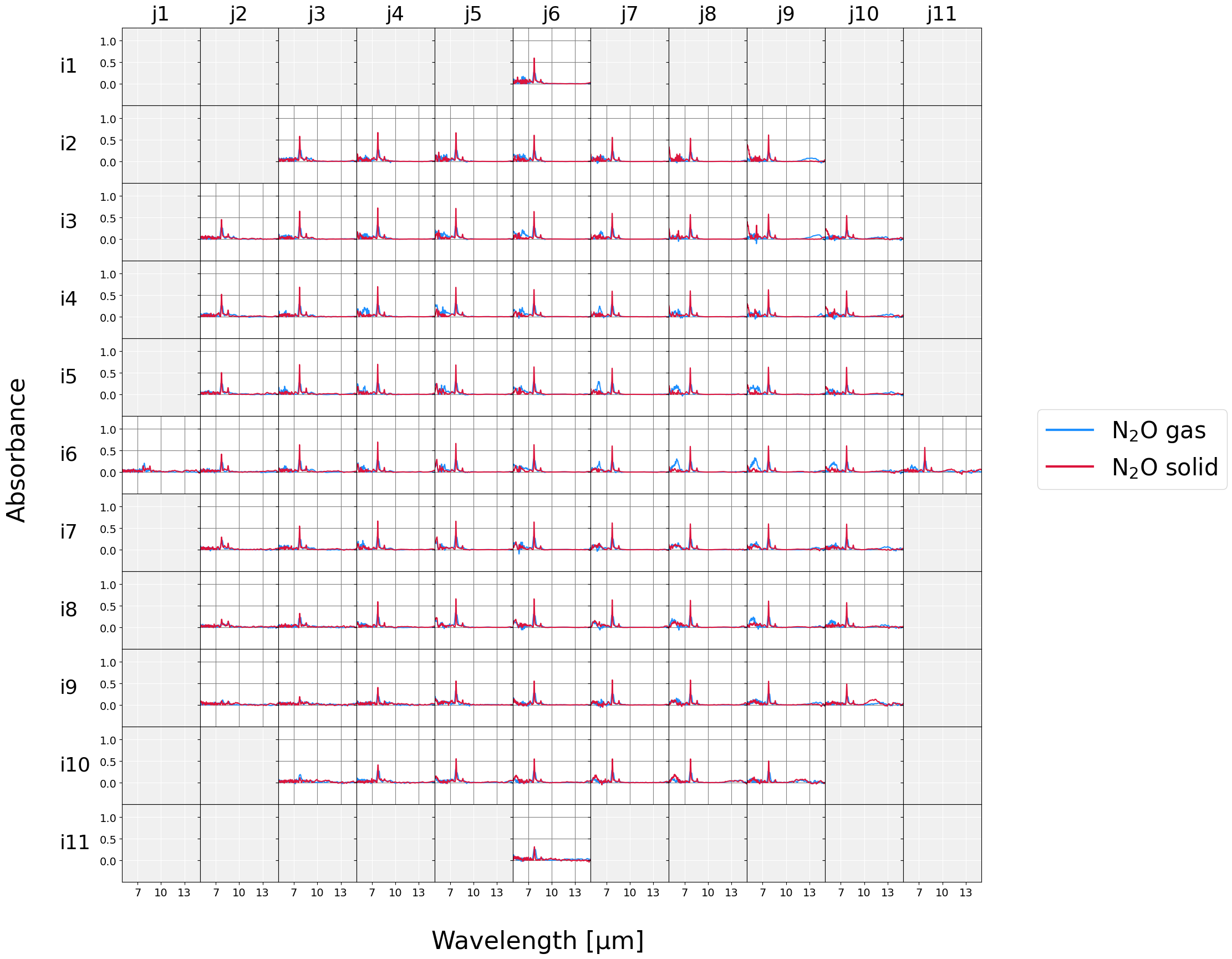}
  \caption{11$\times$11 grid two-dimensional transmission absorption spectra when identical pulses of N$_2$O gas were injected into the vacuum chamber under 107 K cooling conditions (red) and at 302.4 K room temperature (blue). The horizontal axis of each grid spans 5-15 {\textmu}m and the vertical axis represents absorbance from 0.5-1.5.}
  \label{fig:solid_gas}
\end{figure}

\begin{figure}[htbp]
  \centering
  \includegraphics[width=0.8\textwidth]{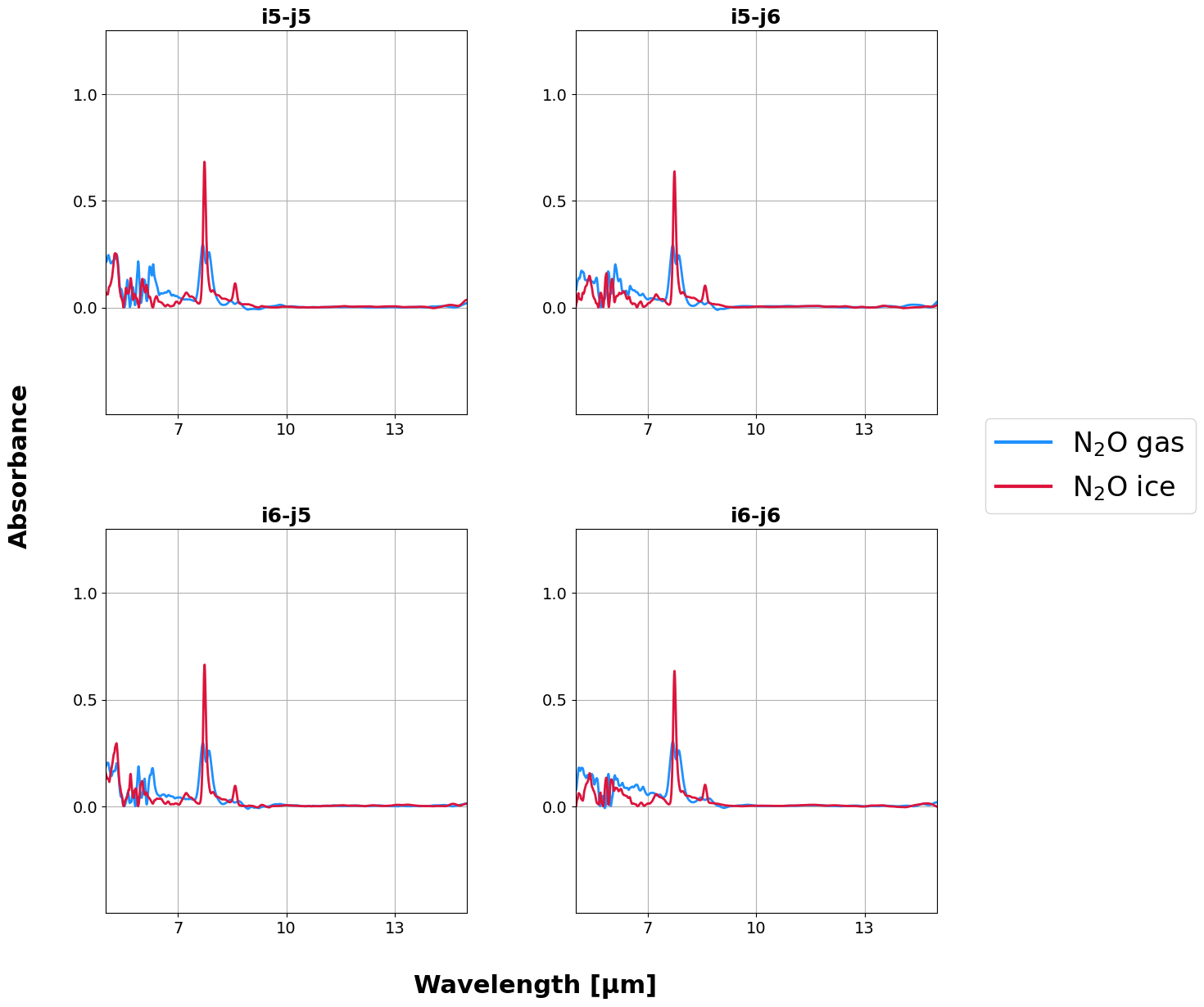}
  \caption{Enlarged transmission absorption spectra of 4 grids near the center of the N$_2$O condensed sample from Figure 3.}
  \label{fig:solid_gas_expansion}
\end{figure}

\clearpage
\subsection{Annealing of N\texorpdfstring{$_2$}{2}O ice}
\ Figure 5 shows the two-dimensional transmission absorption spectra at 23.5, 31.5, 38.5, 40.6, 45, 49.5, 52.8, and 61.5 minutes after N$_2$O gas injection. The pressure is 230 Pa, and the temperatures are 110 K, 115 K, 120 K, 122 K, 115 K, 110 K, 107 K, and 104 K. Figure 6 shows the transmission absorption spectra of four grids near the center of the N$_2$O condensation sample. In all the grids, the torsion (7.27 {\textmu}m), $\nu_1$ (7.75 {\textmu}m), and 2$\nu_2$ (8.60 {\textmu}m) vibration modes of N$_2$O molecules in the condensed phase were enhanced by increasing and decreasing the temperature. In addition, we discovered new peaks at around 12 {\textmu}m and 14 {\textmu}m. The external factors of temperature rise and fall suggest that a phase transition from the amorphous phase to the crystalline phase is occurring in the solid structure of N$_2$O ice, that long-range order is being formed between N$_2$O molecules, and that the density is increasing. In addition, the fact that the absorbance is growing regardless of whether the temperature is rising or falling shows that the crystalline structure changes irreversibly with temperature.

\begin{figure}[htbp]
  \centering
  \includegraphics[width=1.0\textwidth]{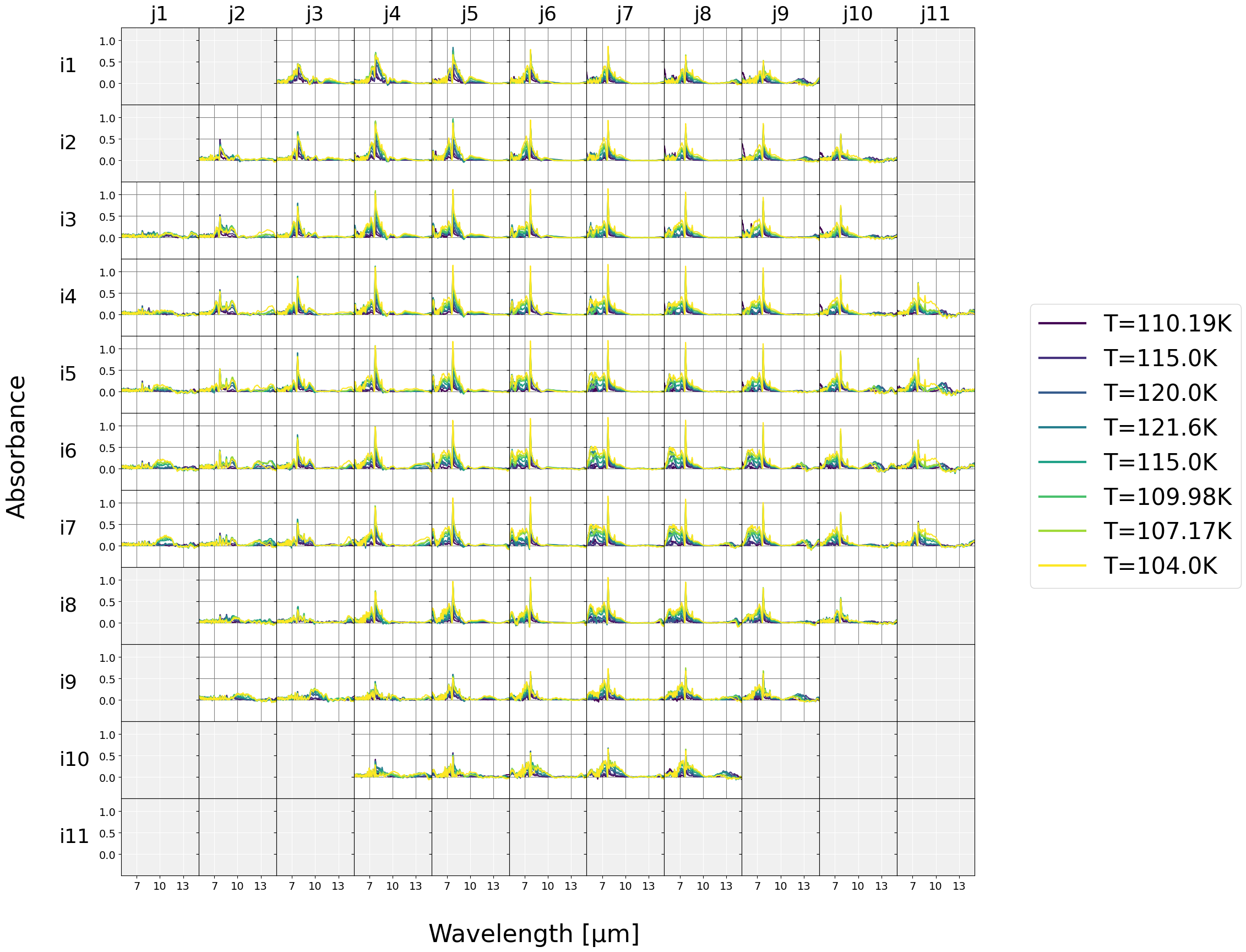}
  \caption{Two-dimensional transmission absorption spectra recorded at temperatures of 110.19 K, 115.0 K, 120.0 K, 121.6 K, 115.0 K, 109.98 K, 107.17 K, and 104.0 K, corresponding to 23.5, 31.5, 38.5, 40.6, 45, 49.5, 52.8, and 61.5 minutes after N$_2$O gas injection.}
  \label{fig:Annealing}
\end{figure}

\begin{figure}[htbp]
  \centering
  \includegraphics[width=1.0\textwidth]{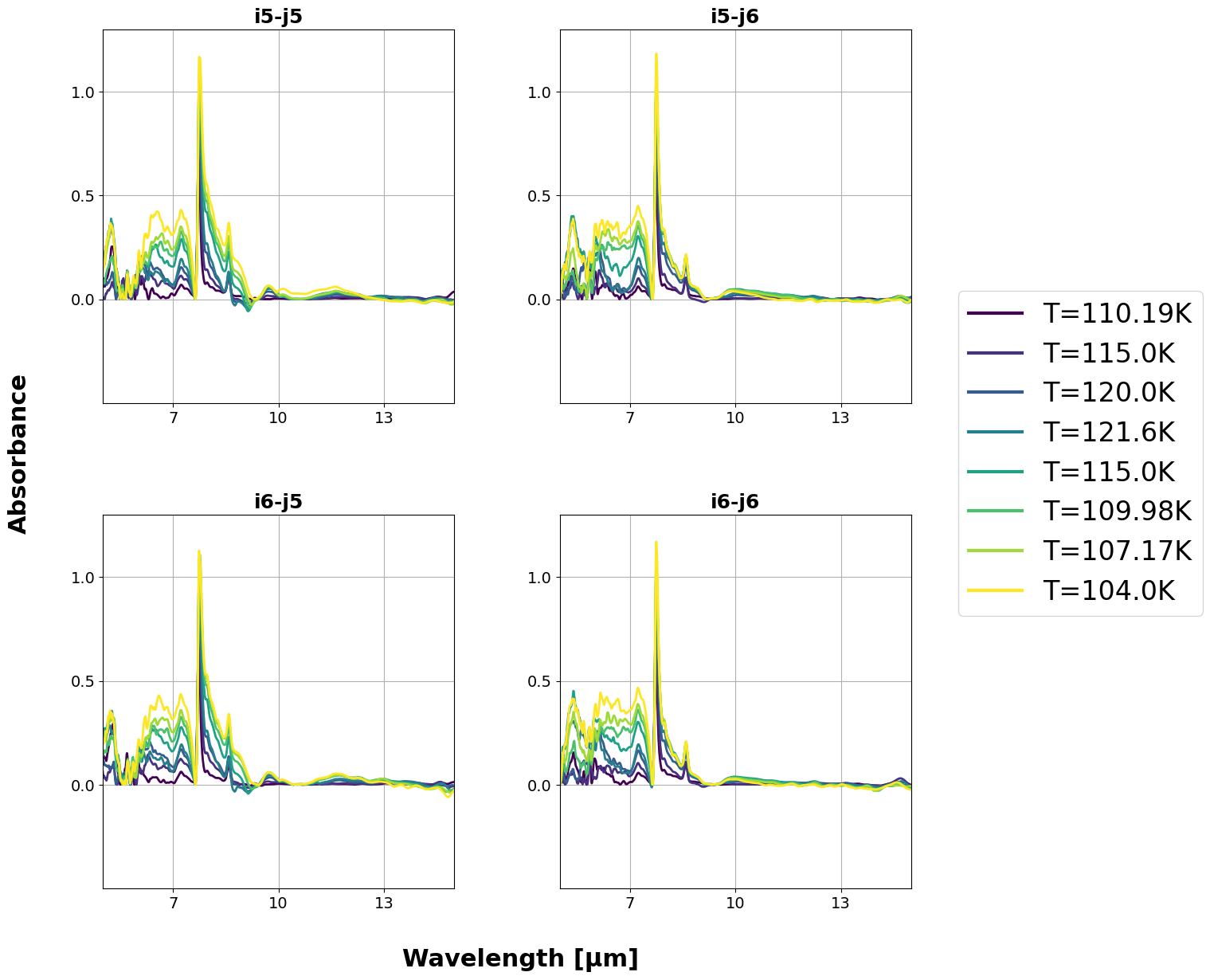}
  \caption{Enlarged two-dimensional transmission absorption spectra of 4 grids near the center of the N$_2$O condensed sample from Figure 6.}
  \label{fig:annealing_expansion}
\end{figure}

\clearpage
\subsection{UV irradiation to \texorpdfstring{N$_2$O}{N2O} ice}
\ Figure 7 presents two-dimensional transmission absorption difference spectra at 0, 10, 20, 30, 40, 50, 60, 70, 80, 90, 120, 180, 240, 300, 360, 420, and 510 minutes after UV irradiation of N$_2$O ice under conditions of 41 Pa pressure and 107 K temperature. These difference spectra represent the subtraction of before-UV irradiation spectra from after-UV irradiation spectra. Figure 8 shows an enlarged view of 4 grids near the center of the N$_2$O condensed sample. Results from HITRAN simulation absorption spectra of N$_2$O, NO, NO$_2$, and O$_3$ under identical temperature and pressure conditions are also presented.

From 20 to 60 minutes of irradiation time, all vibrational modes of N$_2$O increased. Between 60 and 120 minutes, all N$_2$O vibrational modes decreased, while absorbance in the 9$\sim$10 $\mu$m and 11$\sim$14 $\mu$m wavelength ranges initially increased sharply before decreasing. Notably, between 60 and 80 minutes, vibrational modes of N$_2$O$_3$ ($\nu_4$) at 12.73$\sim$12.9 $\mu$m, NO$_2$ ($\nu_2$) and N$_2$O$_4$ ($\nu_{12}$) at 13.28 $\mu$m, and N$_2$O$_5$ ($\nu_{11}$) at 12.58 $\mu$m increased before subsequently decreasing. After 120 minutes, absorption intensities of various nitrogen oxide ice vibrational modes fluctuated at wavelengths 5.33$\sim$5.82 $\mu$m, 6.23$\sim$6.90 $\mu$m, 7.66 $\mu$m, 9.59 $\mu$m and 10.3 $\mu$m, 10.48 $\mu$m, and 14.18 $\mu$m, corresponding to NO($\nu_1$), N$_2$O$_4$($\nu_7$/$\nu_5$), cis-(NO)$_2$, N$_2$O$_3$($\nu_4$) and NO$_2$($\nu_1$), N$_2$O$_3$, NO$_3$, and O$_3$.

\ These temporal changes in absorbance were confirmed to be reproducible when N$_2$O ice was irradiated with UV under identical temperature and pressure conditions. Additionally, differences observed between the central and peripheral regions of the condensed sample suggest that the photodissociation process of N$_2$O ice depends on deposition thickness and crystal structure, indicating complex low-temperature solid-state chemical reactions. Table 3 presents various solid-state vibrational modes confirmed in this study. In contrast, in previous studies, the appendix table from 5 to 7 shows vibrational modes confirmed under multiple conditions.

\begin{table}[ht]
\centering
\captionsetup{justification=centering}
    \caption{N$_x$O$_y$ Absorption band (After UV irradiation to N$_2$O ice at 107 K in condensed phase at temperature 107 K and pressure 41 Pa)}
    \begin{tabular}{p{5cm}p{5cm}p{4cm}}
        \hline
        Chemical species & Wavelength/\textmu m & Vibrational mode \\
        \hline
        NO & 5.34 & $\nu_1$ \\
        N$_2$O$_4$ & 5.33 & N=O str \\
        cis-(NO)$_2$ & 5.37 & N-O s-str (a$_g$(1)) \\
        N$_2$O$_3$ & 5.45 & N=O str \\
        N$_2$O$_2$/cis-(NO)$_2$ & 5.67 & N-O a-str (b$_u$(5)) \\
        N$_2$O$_4$ & 5.82 & $\nu_7$/$\nu_5$ \\
        NO$_2$ & 6.23 & N-O a-str \\
        N$_2$O$_3$ & 6.23 & N-O a-str \\
        N$_2$O$_3$ & 6.31 & $\nu_2$ \\
        NO$_3$ & 6.90 & - \\
        N$_2$O$_3$ & 7.66 & $\nu_3$ \\
        NO$_2$ & 7.66 & $\nu_1$ \\
        O$_3$ & 9.59 & $\nu_3$ O-O a-str \\
        O$_3$ & 10.3 & O-O a-str \\
        NO$_3$ & 10.48 & $\nu_1$ \\
        NO$_3$ & 12.40 & - \\
        N$_2$O$_3$ & 12.73 & $\nu_4$ \\
        N$_2$O$_3$ & 12.90 & NO$_2$ deform \\
        NO$_2$/N$_2$O$_4$ & 13.28 & $\nu_2$/$\nu_{12}$ \\
        N$_2$O$_5$ & 13.58 & $\nu_{11}$ \\
        O$_3$ & 14.18 & - \\
        \hline
    \end{tabular}
    \label{tab:4}
\end{table}

\begin{figure}[htbp]
  \centering
  \includegraphics[width=1.0\textwidth]{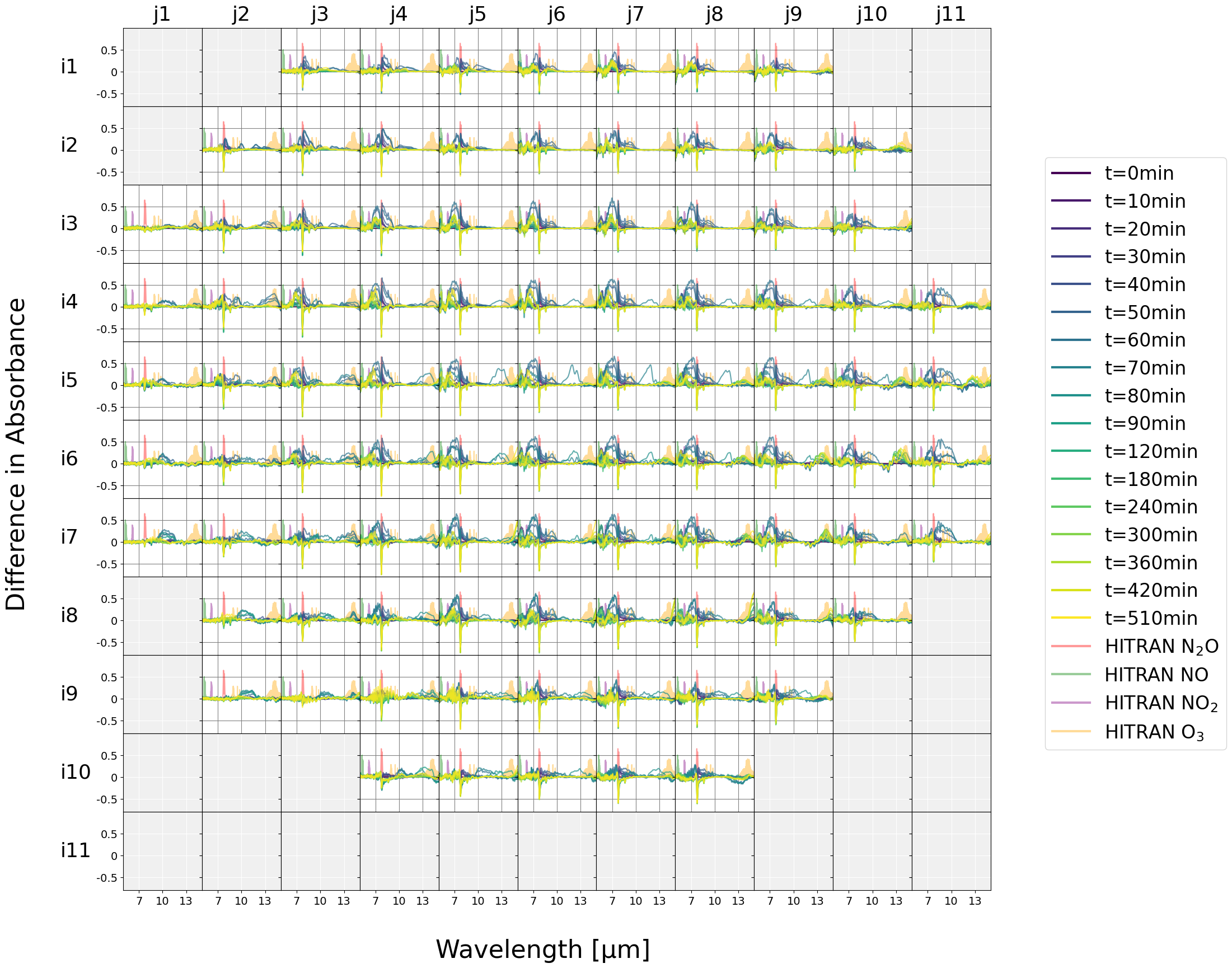}
  \caption{Two-dimensional transmission absorption difference spectra of N$_2$O ice under UV irradiation at a constant temperature of 107 K, measured at 0, 10, 20, 30, 40, 50, 60, 70, 80, 90, 120, 180, 240, 300, 360, 420, and 510 minutes after initial exposure. HITRAN gas simulations for N$_2$O, NO, NO$_2$, and O$_3$ are also shown.}
  \label{fig:UV_difference}
\end{figure}

\begin{figure}[htbp]
  \centering
  \includegraphics[width=1.0\textwidth]{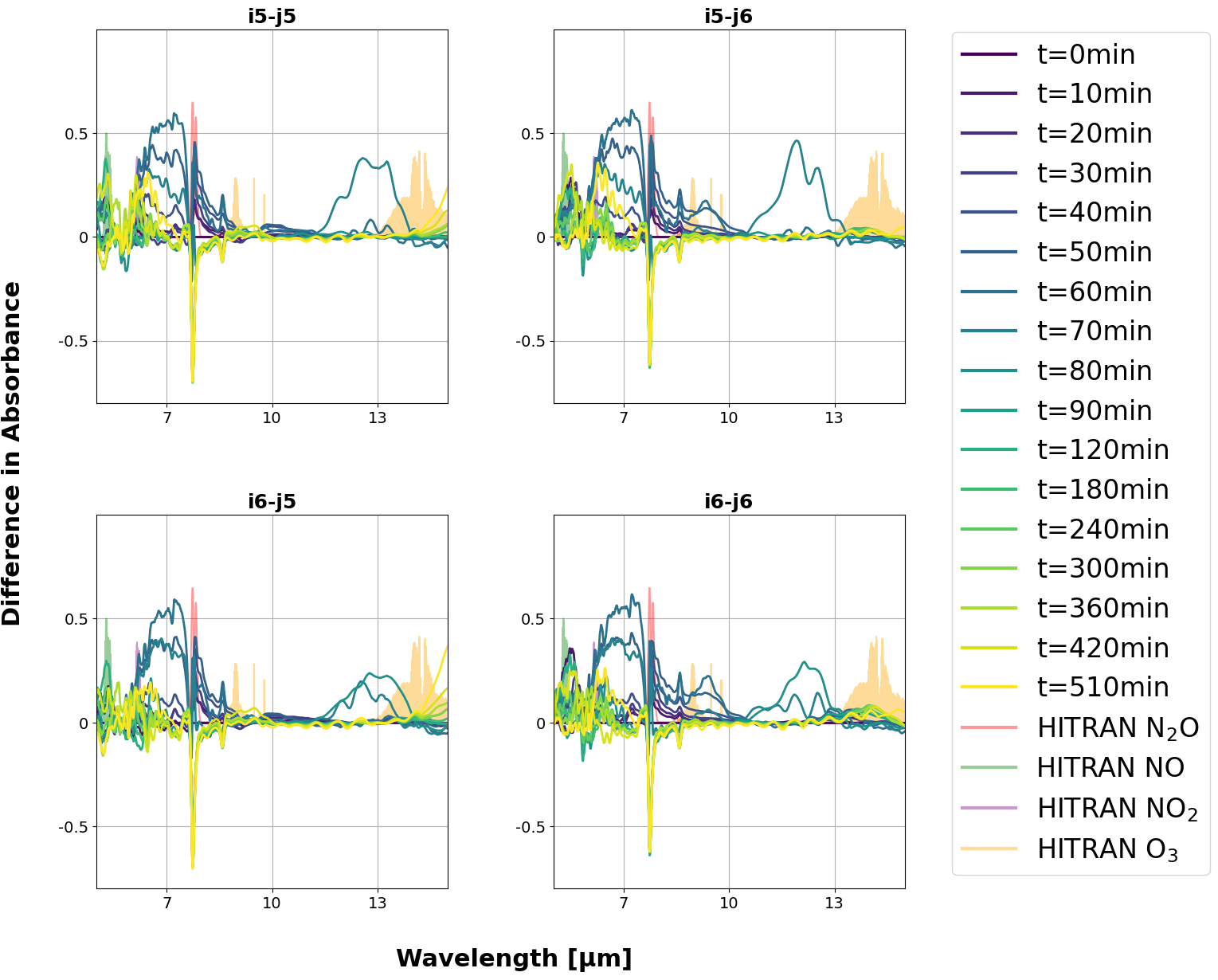}
  \caption{Enlarged two-dimensional transmission absorption difference spectra of 4 grids near the center of the N$_2$O condensed sample from Figure 8.}
  \label{fig:UV_difference_expansion}
\end{figure}

\clearpage
\section{Discussion}
\subsection{Crystal structure and intermolecular distance of N\texorpdfstring{$_2$}{N2O} molecules in the condensed phase}
\ N$_2$O crystallizes in a cubic system with Pa3 space group at temperatures below 182.35 K, containing four N$_2$O molecules per unit cell (Figure 9) (Kuchta and Etters \protect\hyperlink{cite.Kuchta1992}{1992}). The N-N bond distance measures 1.145 \AA, the N-O bond distance is 1.120 \AA, and the permanent dipole moment equals $\mu = 0.17$ D (Wang et al. \protect\hyperlink{cite.Wang2001}{2001}). However, since the intermolecular distances in solid-phase N$_2$O ice remain undetermined, we estimate these distances from the absorbance values of N$_2$O ice obtained in our experiment. For this estimation, we assume that all unit cells of the deposited N$_2$O ice exhibit a cubic crystal structure and form continuous arrangements.

\ The Lambert-Beer law applies since we observe infrared light that has undergone multiple reflections by particles within the N$_2$O ice (Equation 2).

\begin{equation}
A = \varepsilon cl \tag{2}
\label{eq:lambert-beer}
\end{equation}

Define as absorbance $A = 0.7224$, $\varepsilon$ represents the absorption coefficient at 70 K which equals $9.747 \times 10^{-16}$ m/molecule (Hudson et al. \protect\hyperlink{cite.Hudson2017}{2017}), $c$ denote the column density in molecule/cm$^2$ and $l$ indicate the optical path length in meters, which corresponds to the N$_2$O ice thickness. Assuming the ideal gas equation of state, the thickness of N$_2$O ice $l$ in meters can be expressed by equation (3) from equation (2).

\begin{equation}
l = \left[\sum_{i}^{n} \frac{(P_\text{after} - P_\text{before})V_\text{chamber}}{RT_i}\right]\frac{V_\text{chamber}}{S_\text{sh}\rho_\text{crystalline}} \tag{3}
\end{equation}

$S_\text{sh}$ represents the sample holder surface area of $6.907 \times 10^{-3}$ m$^2$, and $\rho_\text{crystalline}$ denotes the density of crystalline N$_2$O ice at 70 K, which equals $1.591 \times 10^3$ kg/m$^3$ (Hudson et al. \protect\hyperlink{cite.Hudson2017}{2017}). $V_\text{chamber}$ indicates the vacuum chamber volume of $2.623 \times 10^{-4}$ m$^3$, $R$ is the gas constant 8.314 J/K mol, $T_i$ represents the temperature in K at each time when N$_2$O gas was ejected through the pulse nozzle, $P_\text{after}$ denotes the pressure in Pa inside the vacuum chamber after each N$_2$O gas ejection, and $P_\text{before}$ indicates the pressure in Pa inside the vacuum chamber before each N$_2$O gas ejection.

The thickness of the N$_2$O ice layer, calculated from Equation (3), is $l = 1.1 \times 10^{-7}$ m, and the column density $c$ is $6.7 \times 10^{21}$ molecules/m$^2$. The number of N$_2$O ice molecules per unit volume, calculated by dividing the column density by the thickness, equals $6.1 \times 10^{28}$ molecule/m$^3$. The N$_2$O ice volume, obtained by multiplying the thickness by the sample holder area, equals $7.6 \times 10^{-10}$ m$^3$. Therefore, the total number of molecules in the deposited N$_2$O ice, calculated by multiplying the number of molecules per unit volume by the N$_2$O ice volume, equals $4.7 \times 10^{19}$ molecules.

Since each unit cell contains four N$_2$O molecules, the number of lattices $k$ equals $1.2 \times 10^{19}$. Consequently, the volume per unit cell, calculated by dividing the N$_2$O ice volume by the number of lattices, equals $6.5 \times 10^{-29}$ m$^3$/molecule. From these calculations, assuming that the length of each side of the cubic crystal system is the same, the intermolecular distance of N$_2$O molecules in the condensed phase is given by Equation (4). Therefore, we estimate that the intermolecular distance of N$_2$O in the condensed phase is approximately 3.5 times the interatomic bond distance.

\begin{equation} 
(6.53 \times 10^{-29})^{\frac{1}{3}} = 4.03 \text{\AA} \tag{4} 
\end{equation}

\begin{figure}[htbp]
  \centering
  \includegraphics[width=0.5\textwidth]{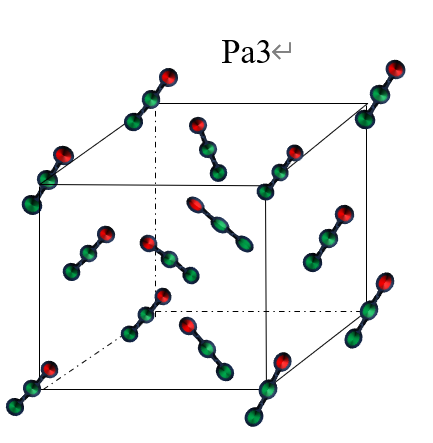}
  \caption{Crystal structure of N$_2$O molecules in the condensed phase. The structure is cubic with Pa3 space group, containing 4 N$_2$O molecules per unit cell (Kuchta and Etters \protect\hyperlink{cite.Kuchta1992}{1992}). Red represents nitrogen atoms, and green represents oxygen atoms.}
  \label{fig:crystal}
\end{figure}

\clearpage
\subsection{Spectral evolution of N\texorpdfstring{$_2$}{2}O ice following thermal annealing}
\ Figure 10 illustrates changes in column density relative to temperature variations for N$_2$O ice vibrational modes: torsion (7.27 {\textmu}m), $\nu_1$ (7.75 {\textmu}m), 2$\nu_2$ (8.60 {\textmu}m), 12 {\textmu}m, and 14 {\textmu}m. In each 11$\times$11 grid, the horizontal axis represents temperature in K, while the vertical axis indicates column density in molecules/cm$^2$ calculated using equations (2) and (3). Arrows within each grid correspond to the sequence of temperature changes.

All vibrational modes exhibit irreversible rate increases with temperature. During the temperature rise period (110.19 K → 121.6 K), the rate of absorbance increase is substantial, whereas during the temperature decrease period (121.6 K → 104.0 K), this rate is smaller. This suggests that crystallization progresses rapidly during temperature rise as intermolecular distances between N$_2$O molecules contract and density increases, while crystallization advances more slowly during temperature decrease. Additionally, near the center of the condensed sample, the increase rates at 12 {\textmu}m and 14 {\textmu}m are smaller but larger at the periphery. This suggests that thinner deposition regions exhibit greater increase rates at 12 {\textmu}m and 14 {\textmu}m.

Figure 11 depicts the rate of column density change $R$ \%/K with temperature for torsion (7.27 {\textmu}m), $\nu_1$ (7.75 {\textmu}m), 2$\nu_2$ (8.60 {\textmu}m), 12 {\textmu}m, and 14 {\textmu}m vibrational modes. The relative change rate $R$ \%/K with temperature is calculated using equation 5:

\begin{equation}
    R = \frac{N_i - N_0}{N_0} \times 100 \tag{5}
\end{equation}

For the heating phase ($T_0$ = 110.19 K → $T_1$ = 121.6 K), we utilized column density values $N_0$ at 110.19 K and $N_i$ at 121.6 K. Conversely, for the cooling phase ($T_0$ = 121.6 K → $T_1$ = 104.0 K), we employed column density values $N_0$ at 121.6 K and $N_i$ at 104.0 K. Figure 11 reveals that during heating, all vibrational modes demonstrate large positive change rates. Notably, torsion and 2$\nu_2$ modes exhibit particularly high positive rates near the center of the condensed sample, indicating faster crystallization in thicker deposition regions. In contrast, outer areas frequently show negative rates, suggesting possible sublimation in thinner deposition regions during heating. These findings demonstrate that N$_2$O ice undergoes irreversible temperature-dependent crystallization exclusively in thicker deposition regions.

\begin{figure}[htbp]
  \centering
  \includegraphics[width=1.0\textwidth]{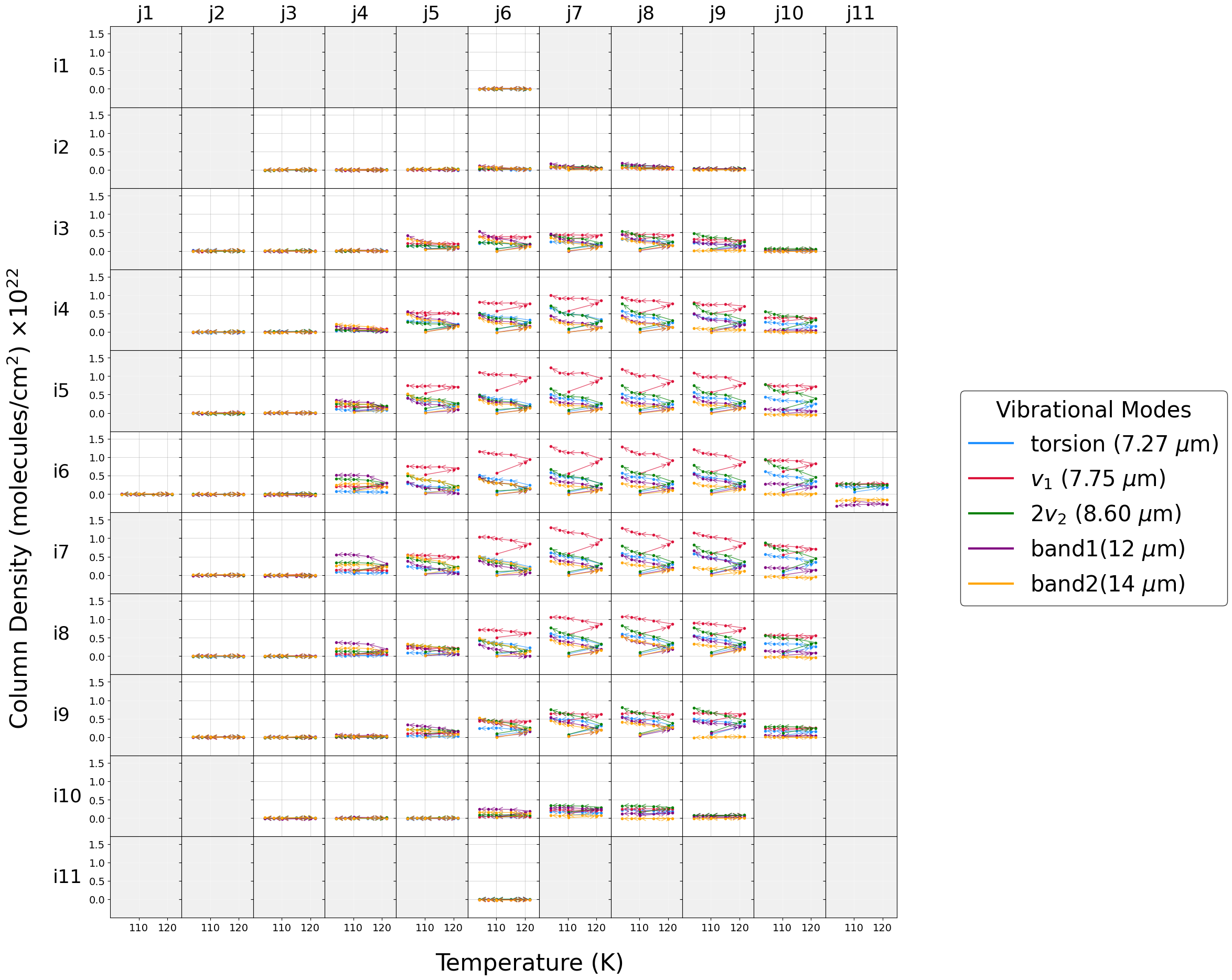}
  \caption{Changes in column density of N$_2$O ice as a function of temperature. The horizontal axis represents temperature in K, while the vertical axis shows column density in molecules/cm$^2$ calculated from equations (2) and (3). Arrows correspond to the sequence of temperature variations.}
  \label{fig:Temp}
\end{figure}

\begin{figure}[htbp]
  \centering
  \includegraphics[width=0.7\textwidth]{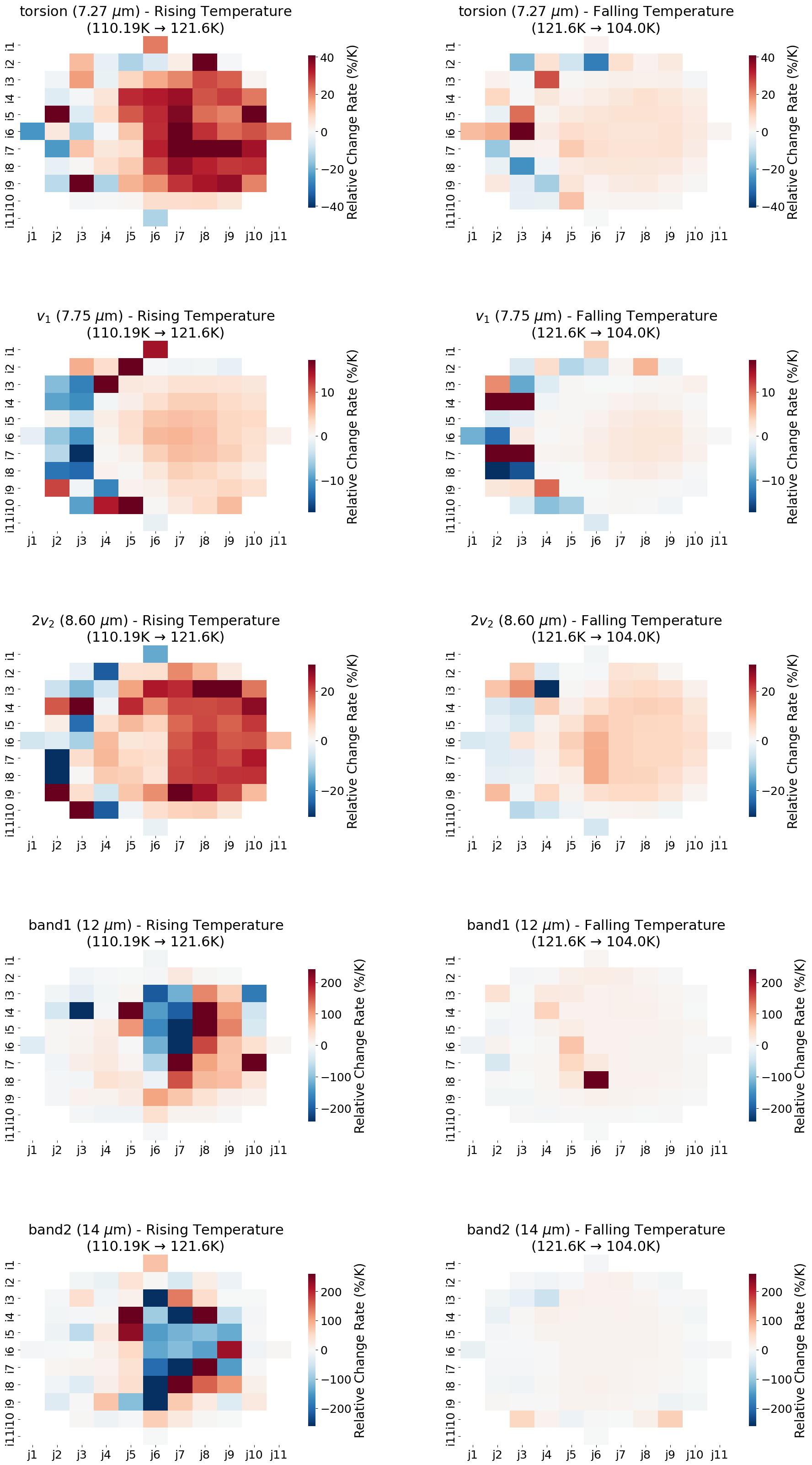}
  \caption{The rate of change in column density for temperature ($\%$/K) for the torsion (7.27 {\textmu}m), $\nu_1$ (7.75 {\textmu}m), 2$\nu_2$ (8.60 {\textmu}m), 12 {\textmu}m, and 14 {\textmu}m vibrational modes during temperature increase (110.19~K → 121.6~K) and temperature decrease (121.6~K → 110.19~K), as determined by equation (5).}
  \label{fig:Temp_rate}
\end{figure}

\clearpage
\subsection{Solid surface chemical reactions of \texorpdfstring{N$_2$O}{N2O} ice by UV irradiation}
\ Figure 12 presents a two-dimensional representation of temporal changes in peak absorbance for each vibrational mode listed in Table 3, displayed as 11×11 grids. The horizontal axis represents UV irradiation time from 0 to 510 min, while the vertical axis indicates column density in molecules/cm$^2$. Experimental data points are fitted using linear interpolation. After UV irradiation, the column density of the $\nu_1$ vibration mode of N$_2$O molecules in the condensed phase at a wavelength of 7.75 {\textmu}m decreases rapidly between 60 and 90 min, suggesting that N$_2$O is photodissociated. Since the dissociation limits of the N-O bond and N-N bond are 741.5 nm and 251.5 nm, respectively, we can assume that these bonds were broken by UV irradiation in the wavelength range of 190 nm to 340 nm. The fact that electrons in N$_2$O molecules transition from the ground state to the excited state and that charge transfer occurs between N$_2$O molecules suggests that a chemical reaction occurs on the surface of the solid. As a result, the formation and dissociation of various nitrogen oxide ices such as N$_2$O$_3$ (m.p. 173 K), NO$_2$ (m.p. 262 K), N$_2$O$_4$ (m.p. 295 K), N$_2$O$_5$ (m.p. 308 K), NO (m.p. 109.5 K), and other nitrogen oxide ice formation and dissociation occurred.

\ We estimate the two-dimensional rate constant $k$ /s for a pseudo-first-order reaction represented by equation (6). We define the formation rate as $k$ /s and the dissociation rate as $-k$ /s, with $[A]$ and $[B]$ representing reactant column densities in molecules/cm$^2$ and $[C]$ denoting product column density in molecules/cm$^2$.

\begin{equation}
    \ce{A + B <=>[$k$][$-k$] C} \tag{6}
\end{equation}

Under this assumption, Figure 12 displays color maps of the normalized rate $|k|$, calculated by dividing the rate of change in column density $[N]$ molecules/cm$^2$ by the column density itself, as expressed in equation (7). Figure 13 illustrates the formation rate, dissociation rate, and their difference.

\begin{equation}
    \text{$|k|$} = \frac{d[N]}{dt} \cdot \frac{1}{[N]}  \tag{7}
\end{equation}

The analysis includes various molecular vibrational modes: NO molecules at 5.34 $\mu$m ($\nu_1$ mode), cis-(NO)$_2$ at 5.67 $\mu$m, NO$_2$ at 6.23 $\mu$m ($\nu_1$ mode), N$_2$O at 7.75 $\mu$m ($\nu_1$ mode), O$_3$ at 14.18 $\mu$m, N$_2$O$_4$ at 13.28 $\mu$m ($\nu_7$/$\nu_5$ modes), N$_2$O$_3$ at 7.66 $\mu$m ($\nu_4$ mode), NO$_3$ at 6.90 $\mu$m, and N$_2$O$_5$ at 13.58 $\mu$m.

Notably, the N$_2$O molecules at 7.75 $\mu$m ($\nu_1$ mode) exhibit a high dissociation rate ($|k|$ $\sim$ $7 \times 10^{-2} \ /\mathrm{s}$) near the condensed sample center, while showing elevated formation rates ($|k|$ $\sim$ $3 \times 10^{-3} \ /\mathrm{s}$) in peripheral regions. We also calculated the D$_2$ lamp photon flux $\phi_{\text{D}_2}$ using equation (8), where $P$ represents the irradiation power $P$ = 6009 $\text{W}/\text{m}^2$, $t$ the irradiation time, $h$ Planck's constant, $c$ the speed of light, and $A$ the sample holder area.

\begin{equation}
\phi_{\text{D}_2} = \int_{190~\text{nm}}^{340~\text{nm}} \frac{Pt\lambda}{hcA} \,d\lambda \tag{8}
\end{equation}

The two-dimensional pattern of N$_2$O molecules at 7.75 $\mu$m ($\nu_1$ mode) suggests that dissociation reactions are dominant in regions with high D$_2$ lamp irradiation intensity ($\phi_{\text{D}_2}$ $\sim 10^{23}$ photons/m$^2$). On the other hand, formation reactions are dominant in regions with low irradiation intensity. For comparison, the formation constant of N$_2$O ice when irradiating a mixture of N$_2$ and CO ice with 10 keV electrons at 10 K was estimated at $6.4 \times 10^{-2}$ /s (Jamieson et al. \protect\hyperlink{cite.Jamieson2005}{2005}).

\ Figure 14 illustrates the chemical reaction network initiated by N$_2$O ice photodissociation, as observed through changes in transmission absorption spectra following UV irradiation. When external UV energy irradiates N$_2$O ice, N-N and N-O bonds break, transforming into N$_2$ or NO within 60-90 minutes (equations 9, 10). Simultaneously, charge transfer occurs at the solid surface, forming oxygen atoms, oxygen molecules, and ozone (equations 11, 12), while N$_2$O$_3$ and N$_2$O$_4$ undergo rapid formation and dissociation between 60-80 minutes (equations 13, 14). From 90-180 minutes, oxygen atoms mediate repeated formation and dissociation of NO and NO$_2$ (equation 15), followed by N$_2$O$_3$ and N$_2$O$_4$ cycling through NO and NO$_2$ intermediates between 180-400 minutes (equations 16, 17). Between 400-510 minutes, cis-(NO)$_2$ forms and dissociates via NO, NO$_2$, and nitrogen atom intermediates (equations 18, 19). Additionally, N$_2$O regenerates from NO$_2$ dissociation (equation 20), subsequently participating in cis-(NO)$_2$ cycling through NO intermediates (equation 21). Ultimately, N$_2$O$_5$ undergoes formation and dissociation cycles from N$_2$O$_3$, N$_2$O$_4$, N$_2$O, and cis-(NO)$_2$ precursors, mediated by oxygen atoms, molecules, and ozone (equations 22-25). This complex N$_x$O$_y$ solid-state chemical reaction network, initiated by N$_2$O ice photodissociation, exhibits time and deposition thickness dependencies. These findings suggest that N$_2$O ice formed from N$_2$ and CO ices on Pluto's and Triton's surfaces and atmospheres likely experiences similar complex N$_x$O$_y$ species formation and dissociation processes upon UV exposure.

\begin{align}
\text{N}_2\text{O} + h\nu &\rightarrow \text{N}_2 + \text{O} \tag{9} \\ 
\text{N}_2\text{O} + h\nu &\rightarrow \text{NO} + \text{N} \tag{10} \\ 
\text{O} + \text{O} &\rightleftharpoons \text{O}_2 \tag{11} \\ 
\text{O}_2 + \text{O} &\rightleftharpoons \text{O}_3 \tag{12} \\ 
\text{N}_2\text{O} + \text{O}_3 &\rightleftharpoons \text{N}_2\text{O}_4 \tag{13} \\ 
\text{N}_2\text{O} + \text{O}_2 &\rightleftharpoons \text{N}_2\text{O}_3 \tag{14} \\ 
\text{NO} + \text{O} &\rightleftharpoons \text{NO}_2 \tag{15} \\ 
\text{NO}_2 + \text{NO} &\rightleftharpoons \text{N}_2\text{O}_3 \tag{16} \\ 
\text{N}_2\text{O}_3 + \text{NO} &\rightleftharpoons \text{N}_2\text{O}_4 \tag{17} \\ 
\text{NO} + \text{NO} &\rightleftharpoons \text{cis-}(\text{NO})_2 \tag{18} \\ 
\text{NO}_2 + \text{N} &\rightleftharpoons \text{cis-}(\text{NO})_2 \tag{19} \\ 
4\text{NO}_2 &\rightleftharpoons \text{N}_2\text{O} + \text{N}_2\text{O}_4 + \text{O}_3 \tag{20} \\ 
\text{N}_2\text{O} + \text{NO} &\rightleftharpoons \text{cis-}(\text{NO})_2 \tag{21} \\ 
\text{N}_2\text{O}_4 + \text{O} &\rightleftharpoons \text{N}_2\text{O}_5 \tag{22} \\ 
\text{N}_2\text{O}_3 + \text{O}_2 &\rightleftharpoons \text{N}_2\text{O}_5 \tag{23} \\ 
\text{N}_2\text{O} + 2\text{O}_2 &\rightleftharpoons \text{N}_2\text{O}_5 \tag{24} \\ 
\text{cis-}(\text{NO})_2 + \text{O}_3 &\rightleftharpoons \text{N}_2\text{O}_5 \tag{25}
\end{align}

\ In the future, we will attempt to quantitatively predict the N$_x$O$_y$ species in the atmosphere by determining the dissociation and formation constants, reaction barriers, and activation energies, taking into account the temperature dependence, by irradiating N$_2$O ice with UV light of intensity ($\phi_{\text{D}_2}$ $\sim 10^{16} \text{ photons/m}^2$) that simulates the actual atmospheric environment under various temperature conditions. Additionally, we aim to conduct UV irradiation of solid mixtures, including N$_2$ ice and CO ice, and to observe vibrational modes of N$_x$O$_y$ species in the unexplored near-infrared wavelength bands of 2-5 $\mu$m and 15-20 $\mu$m using newly developed 2D-FTIR (Table 4) (Zhao et al. \protect\hyperlink{cite.Zhao2024}{2024}). It is possible to compare it with future observations that will be carried out in the 2040s and 2060s, such as the "Arcanum mission" (McKevitt et al. \protect\hyperlink{cite.McKevitt2024}{2024}) and "2044: A DEEP Space Odyssey" (May et al. \protect\hyperlink{cite.May2025}{2025}).

\begin{table}[ht]
\centering
\captionsetup{justification=centering}
    \caption{Vibration modes of N$_x$O$_y$ species in the near-infrared wavelength range that we aim to observe in the future}
        \begin{tabular}{p{4cm}p{3cm}p{3cm}p{5cm}}
        \hline
        Species & Wavelength/\textmu m & Vibrational mode & Reference \\
        \hline
        N$_2$O (amorphous) & 4.47 & $\nu_3$ & Hudson et al. \protect\hyperlink{cite.Hudson2017}{2017}\\
        N$_2$O (crystalline) & 17.0 & $\nu_2$ & // \\
        N$_2$O (crystalline) & 2.85 & $\nu_1$+$\nu_3$ & // \\
        N$_2$O (crystalline) & 2.96 & $2\nu_2$+$\nu_3$ & // \\
        N$_2$O (solid) & 3.55 & $\nu_2$+$\nu_3$ & // \\
        N$_2$O (solid) & 3.88 & $2\nu_1$ & // \\
        N$_2$O (solid) & 4.05 & $\nu_1$+$2\nu_2$ & // \\
        N$_2$O$_4$ (crystalline) & 3.27 & $\nu_1$+$\nu_9$ & Fulvio et al. \protect\hyperlink{cite.Fulvio2019}{2019} \\
        N$_2$O$_4$ (crystalline) & 2.91 & $\nu_5$+$\nu_9$ & // \\
        cis-(NO)$_2$ (solid) & 2.77 & tentative & Bergantini et al. \protect\hyperlink{cite.Bergantini2022b}{2022b} \\
        NO$_2$ (solid) & 3.44 & $\nu_1$+$\nu_3$ & // \\
        \hline
    \end{tabular}
    \label{tab:5}
\end{table}

\begin{figure}[htbp]
  \centering
  \includegraphics[width=1.0\textwidth]{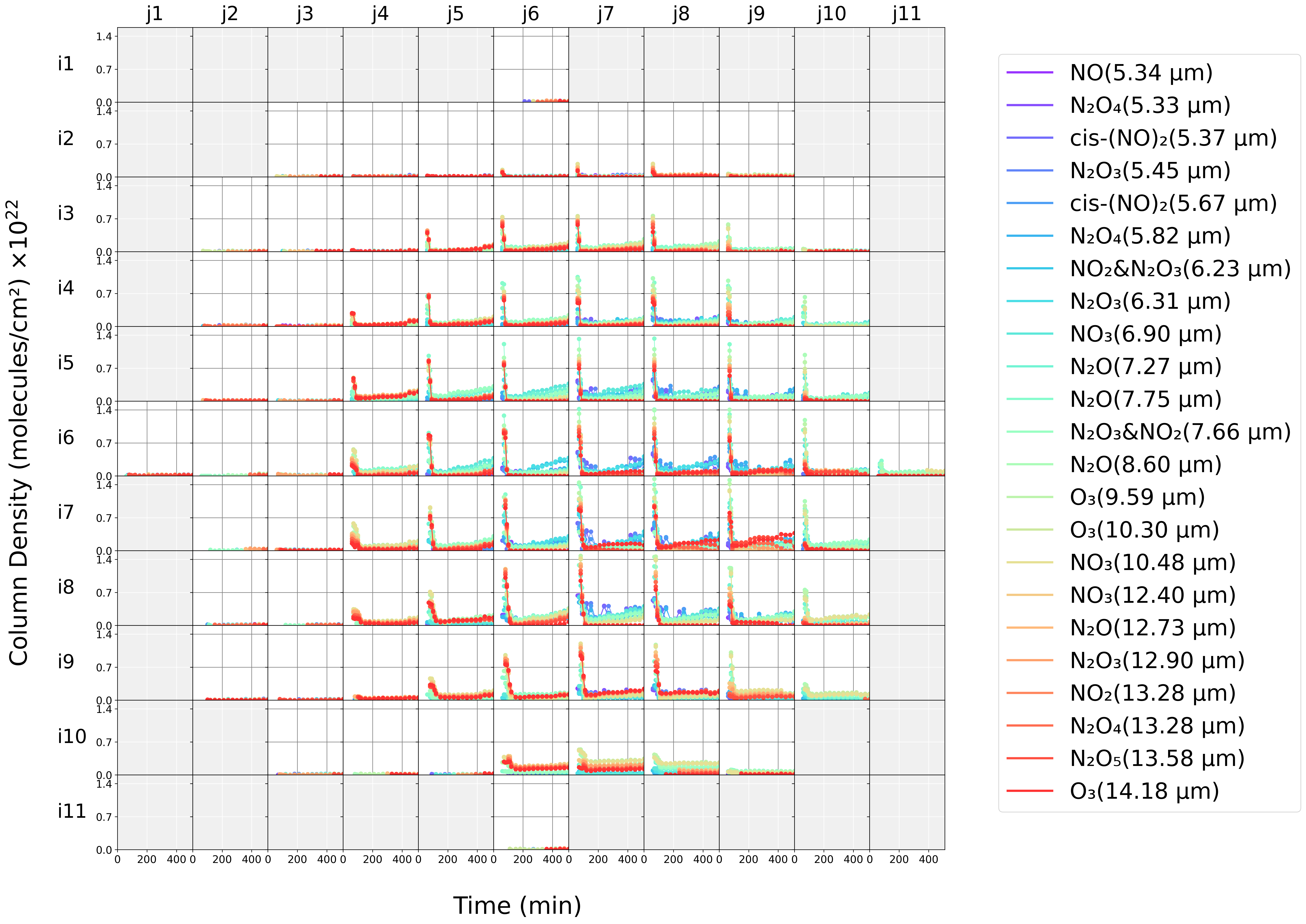}
  \caption{Column density evolution of N$_x$O$_y$ species originating from N$_2$O ice photodissociation as a function of time. The horizontal axis represents UV irradiation time from 0 to 510 min, while the vertical axis shows column density in molecules/cm$^2$ calculated from equations (2) and (3).}
  \label{fig:all_molecules_time_series}
\end{figure}

\begin{figure}[htbp]
  \centering
  \includegraphics[width=1.0\textwidth]{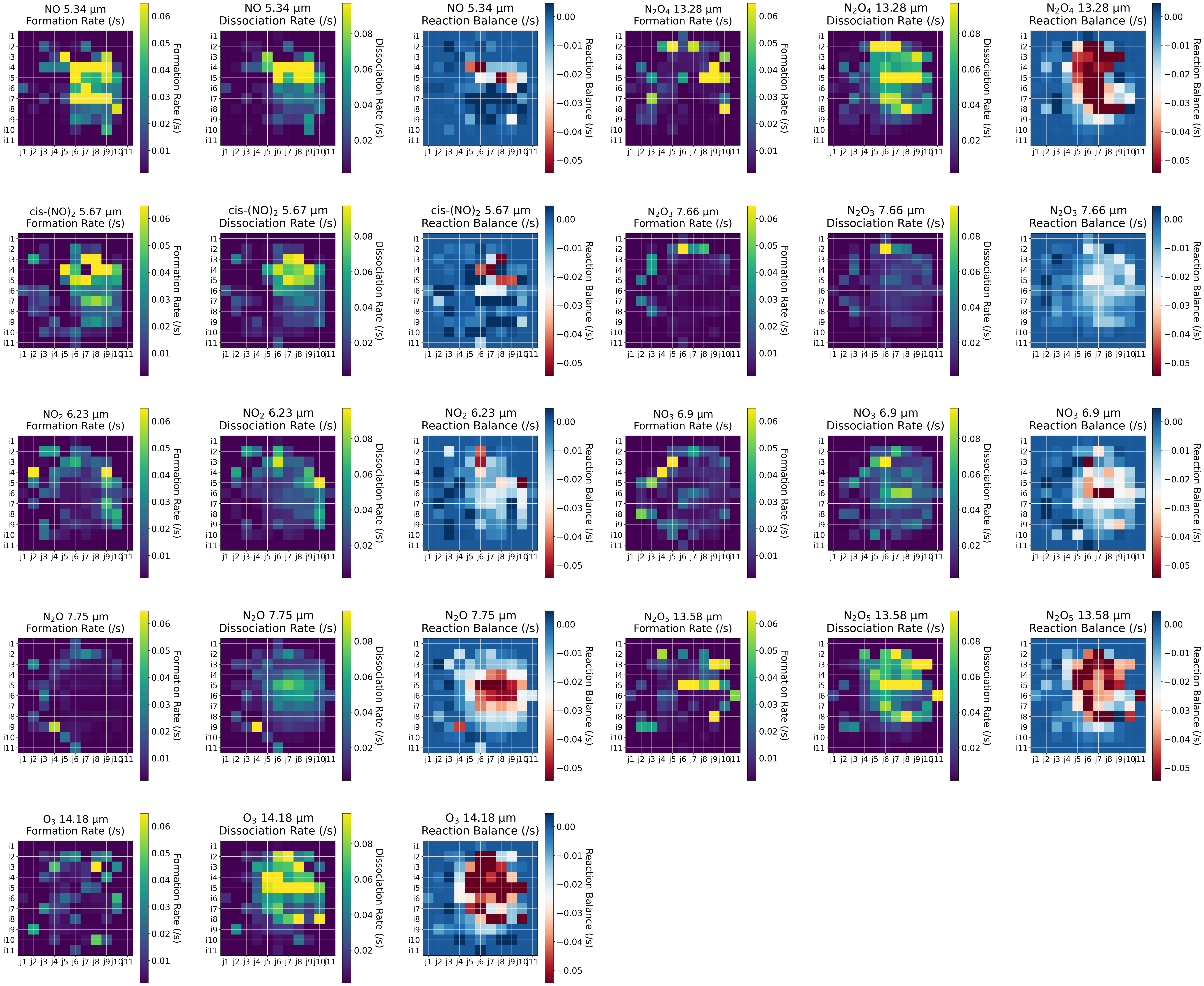}
  \caption{Two-dimensional color map of the normalized rate of change /s calculated by dividing the rate of change in column density for each molecular species, as defined in equation (7).}
  \label{fig:complete_reaction_analysis}
\end{figure}

\begin{figure}[htbp]
  \centering
  \includegraphics[width=1.0\textwidth]{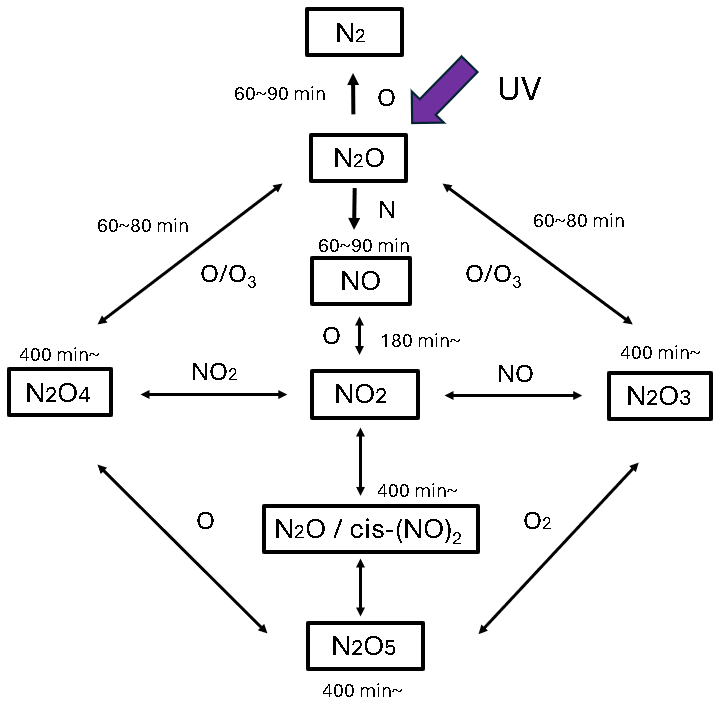}
  \caption{Chemical reaction network of N$_x$O$_y$ ices initiated by photodissociation of N$_2$O ice at 107 K temperature.}
  \label{fig:network}
\end{figure}

\clearpage
\section{Summary}
\ We simulated the atmospheric environment of outer solar system bodies such as Pluto and Triton (100 K, 10$^{-2}$ Pa) and acquired two-dimensional mid-infrared transmission absorption spectra of N$_2$O gas condensates.

\ Spectroscopic imaging revealed a strong absorption band at 7.75 $\mu$m (N$_2$O molecule $\nu_1$ vibrational mode), as well as weak absorption bands at 8.60 $\mu$m (N$_2$O 2$\nu_2$), 7.27 $\mu$m (N$_2$O torsion), and 5.29 $\mu$m (N$_2$O $\nu_1$+$\nu_2$). The absorption peaks of the $\nu_1$ vibrational mode, 2$\nu_2$ vibrational mode, and torsion mode were stronger than those in N$_2$O gas. This suggests that the molecular crystal exhibits strong electrostatic interactions due to significant polarization, and polarized N$_2$O molecules enhance intermolecular interactions via van der Waals forces in the condensed phase.

\ Additionally, we simulated high-temperature regions of 90 K - 110 K by performing annealing (110.19 K → 121.6 K → 104.0 K). As a result, all vibrational modes of N$_2$O molecules increased irreversibly with temperature. During the temperature rise period (110.19 K → 121.6 K), the rate of absorbance increase was substantial, whereas during the temperature decrease period (121.6 K → 104.0 K), this rate was smaller. This suggests that crystallization progresses rapidly during temperature rise and more slowly during temperature decrease.

\ Subsequently, to reproduce photodissociation reactions, we conducted 8.5 hours of ultraviolet irradiation (190 nm-340 nm) using a D$_2$ lamp. Following UV irradiation, all vibrational modes of N$_2$O decreased between 60 and 90 minutes. This likely resulted from breaking N-N bonds (dissociation limit 251.5 nm) and N-O bonds (dissociation limit 741.5 nm) in N$_2$O ice, transforming into N$_2$ or NO. Between 60 and 80 minutes, vibrational modes increased at wavelengths 12.73$\sim$12.9 $\mu$m for N$_2$O$_3$ ($\nu_4$), 13.28 $\mu$m for NO$_2$ ($\nu_2$) and N$_2$O$_4$ ($\nu_{12}$), and 12.58 $\mu$m for N$_2$O$_5$ ($\nu_{11}$). After 120 minutes, absorption intensities of various nitrogen oxide ice vibrational modes fluctuated at wavelengths 5.33$\sim$5.82 $\mu$m, 6.23$\sim$6.90 $\mu$m, 7.66 $\mu$m, 9.59 $\mu$m and 10.3 $\mu$m, 10.48 $\mu$m, and 14.18 $\mu$m, corresponding to NO($\nu_1$), N$_2$O$_4$($\nu_7$/$\nu_5$), cis-(NO)$_2$, N$_2$O$_3$($\nu_4$) and NO$_2$($\nu_1$), N$_2$O$_3$, NO$_3$, and O$_3$. This suggests that UV photons ($\sim 10^{23} \text{ photons/m}^2$) caused electrons in N$_2$O molecules to transition from ground to excited states, generating charge transfer between N$_2$O molecules and triggering solid-surface chemical reactions. Consequently, through the movement of intermediates such as O and species like O$_2$, NO, and NO$_2$, various nitrogen oxides including N$_2$O$_3$ (m.p. 173 K), NO$_2$ (m.p. 262 K), N$_2$O$_4$ (m.p. 295 K), N$_2$O$_5$ (m.p. 308 K), and NO (m.p. 109.5 K) repeatedly formed and dissociated.

\ Based on the decrease and formation rates of peak absorbance for vibrational modes of each molecular species, we estimated pseudo-first-order reaction rates /s in two dimensions. Results showed that dissociation reactions ($7 \times 10^{-3} \ /\mathrm{s}$) proceeded near the center of the condensed sample, while formation reactions ($3 \times 10^{-3} \ /\mathrm{s}$) occurred near the periphery. Therefore, the photodissociation process of N$_2$O ice clearly depends on UV irradiation intensity. This study reveals for the first time that the low-temperature solid-state chemical network of N$_x$O$_y$ species, initiated by photodissociation of N$_2$O ice through UV irradiation, repeatedly undergoes complex dissociation and formation processes depending on time and deposition thickness ($\sim$ 1 {\textmu}m).

\clearpage
\section*{Abbreviations}
IR \hspace{2cm} Infrared

2D FT-MIR \hspace{0.45cm} Imaging Fourier transform mid-infrared spectrometer

ALMA \hspace{1.2cm} The Atacama Large Millimeter/submillimeter Array

UV \hspace{1.75cm} ultraviolet

DEEP \hspace{1.25cm} Discovering Extra-Neptunian and Extrasolar Phenomena

\section*{Declarations}

\subsection*{Availability of data and materials}
\ Raw data were generated at the Department of Earth and Planetary Sciences, Graduate School of Environmental Studies, Nagoya University. Derived data supporting the findings of this study are available from the corresponding author DT and co-authors RK and YH on request. The 2D transmission absorption spectrum creation and FFT transformation codes we developed are available on GitHub (\url{https://github.com/daikitakama}). The specifications for the 2D-FTMIR (NK-0812-TD-NU) can be obtained from the Nisshin Machinery Co. site (\url{https://nissin-kikai.co.jp/company/}). The intensity spectrum of the D$_2$ lamp (A5211TH) used during UV irradiation can also be obtained from the Mitrica Co. site (\url{https://www.milas.co.jp/product_d2.html}).

\subsection*{Competing interests}
\ The authors declare that they have no competing interests.

\subsection*{Funding}
\ This work was supported by JSPS KAKENHI Grant Numbers JP21J00734, JP22K14084, JP20KK0074, JP19H01950, JP21K18640. This work was carried out by the joint research program of the Institute for Space–Earth Environmental Research, Nagoya University.

\subsection*{Authors' contributions}
\ DT summarized the data and wrote the manuscript with RK, SN, and YH contributions. FI contributed to writing the manuscript and supported the experiment. BZ and YL set up the instruments and supported the experiment. RK also developed the FFT data analysis method.

\subsection*{Acknowledgments}
\ The authors thank the Nagoya University Instrument Development Center for creating the vacuum chamber and repairing the dewar bottle.

\subsection*{Authors' information}
\textsuperscript{1}Graduate University for Advanced Studies/Japan Aerospace Exploration Agency (JAXA)/Institute of Space and Astronautical Science (ISAS), Sagamihara, Kanagawa 252-5210, Japan.

\textsuperscript{2}Graduate School of Data Science, Nagoya City University, Nagoya, Aichi 467-8601, Japan.

\textsuperscript{3}Graduate School of Environmental Studies, Nagoya University, Nagoya, Aichi 464-8601, Japan.

\textsuperscript{4}National Institute of Advanced Industrial Science and Technology (AIST), Onogawa 16-1, Tsukuba, Ibaraki 305-8569, Japan.

\clearpage
\section*{Appendix}
\begin{table}[H]
    \centering
    \captionsetup{justification=centering}
    \caption{N$_x$O$_y$ absorption bands from literature (Part 1)}
    \begin{tabular}{p{3cm} p{4.5cm} p{2.5cm} p{2cm} p{2cm}}
        \hline
        \rule{0pt}{1ex}Chemical species & State & Wavelength/\textmu m & Vibrational mode & Reference \\
        \hline
        NO & N$_2$O$_4$/cis-,trans-(NO)$_2$ & 5.29 & $\nu_1$ & Bergantini et al. \protect\hyperlink{cite.Bergantini2022b}{2022b} \\
        N$_2$O$_4$ & N$_2$O ice & 5.35 & $\nu_4$+$\nu_5$/$\nu_1$ & // \\
        N$_2$O$_3$ & N$_2$O ice by $^{136}$Xe$^{23+}$ irradiation & 5.40 & $\nu_1$ & // \\
        N$_2$O$_3$ & N$_2$O ice by $^{136}$Xe$^{23+}$ irradiation & 5.43 & $\nu_1$ & // \\
        N$_2$O$_3$ & N$_2$O ice by $^{136}$Xe$^{23+}$ irradiation & 5.46 & $\nu_1$ & // \\
        N$_2$O$_4$/cis-(NO)$_2$ & N$_2$O ice by $^{136}$Xe$^{23+}$ irradiation & 5.67 & $\nu_9$+R/$\nu_5$ & // \\
        N$_2$O$_4$/trans-(NO)$_2$ & N$_2$O ice by $^{136}$Xe$^{23+}$ irradiation & 5.74 & $\nu_9$/$\nu_5$ & // \\
        N$_2$O$_4$ & N$_2$O ice by $^{136}$Xe$^{23+}$ irradiation & 5.82 & $\nu_7$ & // \\
        NO$_2$ & N$_2$O ice by $^{136}$Xe$^{23+}$ irradiation & 6.20 & $\nu_3$ & // \\
        N$_2$O$_3$ & N$_2$O ice by $^{136}$Xe$^{23+}$ irradiation & 5.91 & - & // \\
        N$_2$O$_3$ & N$_2$O ice by $^{136}$Xe$^{23+}$ irradiation & 6.13 & $\nu_2$ & // \\
        NO$_2$ & N$_2$O ice & 6.21 & asym stretch & // \\
        N$_2$O$_3$ & N$_2$O ice and N$_2$O:H$_2$O mixture by $^{136}$Xe$^{23+}$ irradiation & 6.26 & $\nu_2$ & // \\
        NO$_3$ & N$_2$O:H$_2$O mixture by $^{136}$Xe$^{23+}$ irradiation & 6.99 & tentative &  // \\
        N$_2$O$_3$ & N$_2$O ice and N$_2$O:H$_2$O mixture by $^{136}$Xe$^{23+}$ irradiation & 7.66 & $\nu_3$ & // \\
        N$_2$O & N$_2$O ice & 7.72 & $\nu_1$ & // \\
        N$_2$O & N$_2$O:H$_2$O mixture & 7.76 & $\nu_1$ & // \\
        N$_2$O$_4$ & N$_2$O ice by $^{136}$Xe$^{23+}$ irradiation & 7.92 & $\nu_{11}$ & // \\
        N$_2$O & N$_2$O ice and N$_2$O:H$_2$O mixture & 8.58 & 2$\nu_2$ & // \\
        O$_3$ & N$_2$O ice by $^{136}$Xe$^{23+}$ irradiation & 9.60 & $\nu_3$ & // \\
        O$_3$ & N$_2$O ice by $^{136}$Xe$^{23+}$ irradiation & 9.62 & $\nu_3$ & // \\
        N$_2$O$_3$ & N$_2$O ice by $^{136}$Xe$^{23+}$ irradiation & 12.74 & $\nu_4$ & // \\
        NO$_2$/N$_2$O$_4$ & N$_2$O ice by $^{136}$Xe$^{23+}$ irradiation & 13.28/13.31 & $\nu_2$/$\nu_{12}$ & // \\
        NO & N$_2$O ice by H$^+$ irradiation & 5.35 & $\nu_1$ & Fulvio et al. \protect\hyperlink{cite.Fulvio2019}{2019} \\
        (NO)$_2$ & N$_2$O ice by H$^+$ irradiation & 5.67 & $\nu_5$ & // \\
        (NO)$_2$ & N$_2$O ice by H$^+$ irradiation & 5.35 & $\nu_1$ & // \\
        NO$_2$ & N$_2$O ice by H$^+$ irradiation & 13.28 & $\nu_2$ & // \\
        NO$_3$ & N$_2$O ice by H$^+$ irradiation & 6.20 & $\nu_3$ & // \\
        NO$_3$ & N$_2$O ice by H$^+$ irradiation & 10.53 & $\nu_1$ & // \\
        NO$_3$ & N$_2$O ice by H$^+$ irradiation & 6.99 & - & // \\
        N$_2$O$_3$ & N$_2$O ice by H$^+$ irradiation & 12.74 & $\nu_4$ & // \\
        N$_2$O$_3$ & N$_2$O ice by H$^+$ irradiation & 7.66 & $\nu_3$ & // \\
        N$_2$O$_3$ & N$_2$O ice by H$^+$ irradiation & 6.26 & $\nu_2$ & // \\
        N$_2$O$_3$ & N$_2$O ice by H$^+$ irradiation & 5.46 & $\nu_1$ & // \\
        N$_2$O$_4$ & N$_2$O ice by H$^+$ irradiation & 13.28 & $\nu_{12}$ & // \\
        N$_2$O$_4$ & N$_2$O ice by H$^+$ irradiation & 7.94 & $\nu_{11}$ & // \\
        N$_2$O$_4$ & N$_2$O ice by H$^+$ irradiation & 5.81 & $\nu_7$/$\nu_5$ & // \\
        N$_2$O$_4$ & N$_2$O ice by H$^+$ irradiation & 5.74 & $\nu_9$ & // \\
        N$_2$O$_4$ & N$_2$O ice by H$^+$ irradiation & 5.67 & $\nu_6$+$\nu_{11}$ & // \\
        N$_2$O$_5$ & N$_2$O ice by H$^+$ irradiation & 13.59 & $\nu_{11}$ & // \\
        N$_2$O$_5$ & N$_2$O ice by H$^+$ irradiation & 8.05 & $\nu_{10}$ & // \\
        N$_2$O$_5$ & N$_2$O ice by H$^+$ irradiation & 7.46 & $\nu_2$ & // \\
        N$_2$O$_4$; NO$_2$ & NO$_2$:N$_2$O$_4$ mixture & 13.24$-$13.48 & $\nu_{12}$:$\nu_2$ & // \\
        N$_2$O$_3$ & NO$_2$:N$_2$O$_4$ mixture & 12.74 & $\nu_4$ & // \\
        Unknown & $^{64}$Ni$^{24+}$ irradiation & 7.31 & - & // \\
        \hline
    \end{tabular}
    \label{tab:6}
\end{table}

\begin{table}[H]
    \centering
    \captionsetup{justification=centering}
    \caption{N$_x$O$_y$ absorption bands from literature (Part 2)}
    \begin{tabular}{p{3cm} p{4cm} p{2.5cm} p{2cm} p{3cm}}
        \hline
        \rule{0pt}{1ex}Chemical species & State & Wavelength/\textmu m & Vibrational mode & Reference \\
        \hline
        NO$_3$ & NO$_2$:N$_2$O$_4$ mixture & 12.42 & - & Fulvio et al. \protect\hyperlink{cite.Fulvio2019}{2019} \\
        NO$_3$ & NO$_2$:N$_2$O$_4$ mixture & 10.42$-$10.56 & $\nu_1$ & // \\
        N$_2$O$_4$ & NO$_2$:N$_2$O$_4$ mixture & 7.92$-$7.99 & $\nu_{11}$ & // \\
        NO$_2$; N$_2$O; N$_2$O$_3$ & NO$_2$:N$_2$O$_4$ mixture & 7.67$-$7.65 & $\nu_1$:$\nu_1$:$\nu_3$ & // \\
        NO$_3$ & NO$_2$:N$_2$O$_4$ mixture & 6.88$-$6.93 & - & // \\
        N$_2$O$_3$ & NO$_2$:N$_2$O$_4$ mixture & 6.31 & $\nu_2$ & // \\
        NO$_2$ & NO$_2$:N$_2$O$_4$ mixture & 6.19 & $\nu_3$ & // \\
        N$_2$O$_4$ & NO$_2$:N$_2$O$_4$ mixture & 7.91$-$5.83 & $\nu_7$/$\nu_5$ & // \\
        N$_2$O$_4$ & NO$_2$:N$_2$O$_4$ mixture & 5.74$-$5.77 & $\nu_9$ & // \\
        N$_2$O$_4$; (NO)$_2$ & NO$_2$:N$_2$O$_4$ mixture & 5.67$-$5.77 & $\nu_6$+$\nu_{11}$:$\nu_{15}$ & // \\
        NO; (NO)$_2$ & NO$_2$:N$_2$O$_4$ mixture & 5.34$-$5.35 & $\nu_1$:$\nu_1$ & // \\
        NO & NO$_2$:N$_2$O$_4$ mixture by H$^+$ irradiation & 5.35 & $\nu_1$ & // \\
        (NO)$_2$ & NO$_2$:N$_2$O$_4$ mixture by H$^+$ irradiation & 5.67 & $\nu_5$ & // \\
        (NO)$_2$ & NO$_2$:N$_2$O$_4$ mixture by H$^+$ irradiation & 5.35 & $\nu_1$ & // \\
         N$_2$O & NO$_2$:N$_2$O$_4$ mixture by H$^+$ irradiation & 7.66 & $\nu_1$ & // \\
        N$_2$O$_3$ & NO$_2$:N$_2$O$_4$ mixture by H$^+$ irradiation & 7.66 & $\nu_3$ & // \\
        N$_2$O$_3$ & NO$_2$:N$_2$O$_4$ mixture by H$^+$ irradiation & 6.26 & $\nu_2$ & // \\
        N$_2$O$_3$ & NO$_2$:N$_2$O$_4$ mixture by H$^+$ irradiation & 5.46 & $\nu_1$ & // \\
        N$_2$O$_5$ & NO$_2$:N$_2$O$_4$ mixture by H$^+$ irradiation & 13.59 & $\nu_{11}$ & // \\
        N$_2$O$_5$ & NO$_2$:N$_2$O$_4$ mixture by H$^+$ irradiation & 8.05 & $\nu_{10}$ & // \\
        N$_2$O$_5$ & NO$_2$:N$_2$O$_4$ mixture by H$^+$ irradiation & 7.46 & $\nu_2$ & // \\
        N$_2$O$_3$ & NO$_2$:N$_2$O$_4$ mixture by H$^+$ irradiation & 12.74 & $\nu_4$ & // \\
        NO$_2$ & N$_2$:H$_2$O mixture & 6.20 & - & Barros et al. \protect\hyperlink{cite.deBarros2017}{2017} \\
        NO$_3$ & N$_2$:H$_2$O mixture & 6.82 & - & // \\
        NO & N$_2$:H$_2$O mixture & 5.28 & - & // \\
        O$_3$ & N$_2$:H$_2$O mixture & 9.60 & - & // \\
        N$_2$O$_5$ & N$_2$:H$_2$O mixture & 7.66 & - & // \\
        N$_2$O$_4$ & N$_2$:H$_2$O mixture & 7.67 & - & // \\
        N$_2$O$_3$ & N$_2$:H$_2$O mixture & 5.45 & - & // \\
        N$_2$O$_2$ & N$_2$:H$_2$O mixture & 5.66 & - & // \\
        \hline
    \end{tabular}
    \label{tab:7}
\end{table}

\begin{table}[H]
    \centering
    \captionsetup{justification=centering}
    \caption{N$_x$O$_y$ absorption bands from literature (Part 3)}
    \begin{tabular}{p{3cm} p{3cm} p{2.5cm} p{3cm} p{3cm}}
        \hline
        \rule{0pt}{1ex}Chemical species & State & Wavelength/\textmu m & Vibrational mode & Reference \\
        \hline
        NO & (SiO)$_x$ substrate NO+O$_2$ reaction & 5.27 & - & Minissale et al. \protect\hyperlink{cite.Minissale2014}{2014} \\
        cis-(NO)$_2$ & (SiO)$_x$ substrate NO+O$_2$ reaction & 5.37 & N-O s-str(a$_g$(1)) & // \\
        cis-(NO)$_2$ & (SiO)$_x$ substrate NO+O$_2$ reaction & 5.63 & N-O s-str(b$_u$(5)) & // \\
        NO$_2$ & (SiO)$_x$ substrate NO+O$_2$ reaction & 6.23 & N-O a-str & // \\
        ONNO$_2$ & (SiO)$_x$ substrate NO+O$_2$ reaction & 7.62 & N=O s-str & // \\
        ONNO$_2$ & (SiO)$_x$ substrate NO+O$_2$ reaction & 5.45 & N-O s-str & // \\
        ONNO$_2$ & (SiO)$_x$ substrate NO+O$_2$ reaction & 6.23 & N-O s-str & // \\
        ONNO$_2$ & Gold substrate NO+N reaction & 7.63 & NO$_2$ deform & // \\
        N$_2$O$_4$ & (SiO)$_x$ substrate NO+O$_2$ reaction & 5.33 & N=O str & // \\
        N$_2$O$_4$ & (SiO)$_x$ substrate NO+O reaction & 5.74 & NO$_2$ a-str(b$_3$$_u$(11)) & // \\
        N$_2$O$_4$ & (SiO)$_x$ substrate NO+O reaction & 7.96 & NO$_2$ a-str(b$_2$$_u$(11)) & // \\
        N$_2$O & H$_2$O ice substrate NO+N$_2$O reaction & 7.78 & N=O str & // \\
        $^{16}$O$^{16}$O$^{16}$O$_3$ & (SiO)$_x$ substrate NO+O$_3$ reaction & 9.58 & O-O a-str & // \\
        $^{18}$O$^{18}$O$^{16}$O$_3$ & CO ice substrate NO+N$_2$O reaction & 10.27 & O-O a-str & // \\
        NO & $^{14}$N$^+$ irradiation & 5.37 & NO Monomer stretch & Almeida et al. \protect\hyperlink{cite.Almeida2017}{2017} \\
        N$_2$O$_2$ & $^{14}$N$^+$ irradiation & 5.67 & On-No Antisymmetric stretch & // \\
        N$_2$O$_4$/N$_2$O$_5$ & $^{14}$N$^+$ irradiation & 5.75 & NO stretch & // \\
        N$_2$O$_5$ & $^{14}$N$^+$ irradiation & 5.85 & (B) Antisymmetric NO stretch & // \\
        N$_2$O$_4$ & $^{14}$N$^+$ irradiation & 7.94 & (B$_3$$_u$) NO stretch & // \\
        NO$_2$ & $^{14}$N$^+$ irradiation & 6.20 & Antisymmetric stretch & // \\
        O$_3$ & $^{14}$N$^+$ irradiation & 9.63 & Antisymmetric stretch & // \\
        N$_2$O$_3$ & $^{14}$N$^+$ irradiation & 6.28 & NO$_2$ Antisymmetric stretch & // \\
        N$_2$O$_3$ & $^{14}$N$^+$ irradiation & 7.65 & NO$_2$ Symmetric stretch & // \\
        N$_2$O$_3$ & $^{14}$N$^+$ irradiation & 12.76 & Deformation of NO$_2$ group & // \\
        N$_2$O$_4$ & e$^-$ irradiation & 7.93 & $\nu_{12}$ & Mifsud et al. \protect\hyperlink{cite.Mifsud2022}{2022} \\
        \hline
    \end{tabular}
    \label{tab:8}
\end{table}

\end{document}